%% file: main.tex
\documentclass[a4paper,11pt]{article}
\usepackage{jinstpub}
\usepackage{lineno}
\usepackage{graphicx}
\usepackage{amsmath}
\usepackage{siunitx}
\usepackage{hyperref}
\usepackage{booktabs}
\usepackage{multirow}
\usepackage{subcaption}
\usepackage{xcolor}
\usepackage{orcidlink}
\usepackage{textcomp}
\usepackage{circuitikz}

\usepackage{pgfplots}
\title{\boldmath The RUBIK satellite payload concept: a 3D-segmented plastic scintillation-based particle tracking detector }

\author[a,1]{Roman Bergert \orcidlink{0000-0003-4109-7534},\,
  \note{Corresponding author.}}
\author[a]{Dzmitry Kazlou \orcidlink{0009-0002-2658-2347},\,}
\author[a]{Hartmut Schotte,}
\author[a,b]{Niclas Fiedler \orcidlink{0009-0000-4259-4503},\,}
\author[a]{Hans-Georg Zaunick \orcidlink{0000-0002-6835-0109} }
\author[b,c]{Kilian-Simon Baumann \orcidlink{0000-0003-1223-523X} }
\author[a]{and Kai-Thomas Brinkmann \orcidlink{0000-0002-4387-8193}\,}

\emailAdd{roman.bergert@physik.jlug.de}

\affiliation[a]{2\textsuperscript{nd}~Physics Institute Justus Liebig University,
Giessen, Germany}

\affiliation[b]{Institute of Medical Physics and Radiation Safety University of Applied Sciences, Giessen, Germany}

\affiliation[c]{Marburg Ion-Beam Therapy Center,
Marburg, Germany}

\date{\today}

\abstract{%
We present the design concept and first experimental results of RUBIK,
a three-dimensional resolving plastic scintillation detector developed
as a scientific payload for the Research and Observation in Medium Earth
Orbit satellite mission of the Institute of Space Systems
of the University of Stuttgart.
The name RUBIK is inspired by the Rubik's Cube, a familiar example of
a three-dimensional segmented structure, reflecting the
\(5{\times}5{\times}5\) modular arrangement of the sensitive volume.
The sensitive volume consists of 125 PVT-based plastic scintillator unit
cubes, each penetrated by two perpendicularly aligned
wavelength-shifting fibres, which guide scintillation photons to
silicon photomultipliers (SiPMs), providing 50 independent readout
channels.
Custom readout electronics, comprising two discriminator boards with
per-channel programmable thresholds, an FPGA-based data-acquisition
board, and a dedicated power and bias distribution board, have been
developed exclusively from commercial off-the-shelf components.
A single-layer prototype with ten readout channels was used for concept validation through SiPM bias voltage and temperature characterisation, and a time-over-threshold based amplitude reconstruction method was validated against radioactive source and cosmic ray measurements.
First hit-pattern measurements with \SI{220}{\MeV} protons at a clinical proton therapy facility confirmed that spatially resolved hit clusters consistent with the expected beam geometry were observable in the single-layer readout, demonstrating the capability to resolve particle-induced signal patterns as a first step towards 3D track reconstruction in the full instrument.
}


\begin{document}
\maketitle
\flushbottom

\section{Introduction}
\label{sec:intro}

The measurement of relativistic charged particle fluxes in low and
medium Earth orbit is relevant both for understanding the near-Earth
radiation environment~\cite{baker2018space,zheng2024overview,barmin2014near} and for space-weather
research, as demonstrated by a series of dedicated orbital
experiments~\cite{AMS2015,CALET2019,DAMPE2019,spence2022achievements,adriani2011pamela}.
Beyond radiation monitoring, primary cosmic rays detected in orbit can 
trigger extensive air showers observable by ground-based and deep-ice 
detector arrays~\cite{Auger2015,TA2013,IceCube2013}. An event-by-event 
correlation between such orbital detections and their associated 
ground-level air showers has yet to be demonstrated experimentally.
Existing space-based missions targeting cosmic-ray air showers
approach this from above by detecting atmospheric fluorescence and
Cherenkov light~\cite{khrenov2017first,JEMEUSO2020}, a direct particle-level event-by-event correlation between an
orbital detector and a ground-based secondary air shower observatory has,
to the authors' knowledge, not been demonstrated experimentally.
Establishing such a correlation requires a compact orbital detector
capable of three-dimensional track reconstruction with precise event
timing, operating continuously over the full duration of the mission
in a high-radiation environment.

The Research and Observation in Medium Earth Orbit (ROMEO) mission,
developed at the Institute of Space Systems (IRS) of the University of
Stuttgart~\cite{Loeffler2022}, provides a \SI{70}{kg} satellite
platform specifically designed to qualify new technologies in low and
medium earth orbit.
The mission is structured in three phases: a sun-synchronous orbit (SSO)
at \SI{600}{km} altitude (Phase~1), a manoeuvre raising the apogee to above
\SI{2000}{km} using an on-board water electrolysis propulsion system
(Phase~2, $2400\times337$~km), and a final elliptical orbit
($3026\times314$~km) that traverses the inner Van Allen radiation belt
(Phase~3).
This multi-phase walking-orbit profile is unique in sweeping a wide range
of orbital altitudes within a single one-year mission, creating an
opportunity to map particle rate distributions across the most
radiation-intense region of the near-Earth environment --- a measurement
that is inaccessible from a fixed-altitude platform.

RUBIK --- named after the Rubik's Cube, whose characteristic
\(3{\times}3{\times}3\) layer structure inspired the
\(5{\times}5{\times}5\) arrangement of the sensitive volume --- is the
particle detector payload of the ROMEO mission, developed at Justus
Liebig University (JLU) Giessen/Germany.
A central design requirement of the ROMEO mission is to demonstrate that
commercial off-the-shelf (COTS) components --- without radiation-hardened
or application-specific integrated circuits --- can be used to build a
functional scientific instrument capable of surviving and operating in the
MEO radiation environment.
RUBIK adopts this philosophy throughout: every component in the signal
chain, readout electronics, and power distribution is based on commercially
available parts, and the electronics are realised as a full-custom design
based entirely on COTS devices.
The design is constrained by the satellite mission to a current budget of \SI{2}{A}, a volume envelope of \(1\text{U} = 10\times10\times10\)~\SI{}{cm^3}, and a mass allocation of \SI{1.5}{kg}.
Within these boundaries, per-channel energy measurement via
waveform digitisation --- which would require either a dedicated
ADC per channel or a custom mixed-signal readout ASIC --- is incompatible with the mission requirements.
RUBIK is therefore designed primarily as a tracking detector and particle 
flux rate counter. Its core science observables are the three-dimensional 
hit pattern, reconstructed track direction with a precise event timestamp, and 
per-orbit event rate.

A time-over-threshold (ToT) measurement, implemented at negligible additional hardware cost within the existing discriminator architecture, provides a per-channel proxy for deposited energy. Combined with the five-layer coincidence topology, this enables particle characterisation by energy deposition profile — distinguishing stopped protons from minimum-ionising particles traversing the entire volume — and constitutes a first step towards energy-weighted event classification without a dedicated waveform-digitising stage. This paper presents the full detector concept, the electronics design, the data-acquisition architecture, and first experimental and simulation results obtained with a ($5{\times}5{\times}1$) single-layer prototype.
\section{Scientific Objectives}
\label{sec:science}

\begin{figure}[h!]
  \begin{subfigure}[b]{0.32\textwidth}
    \centering
    \includegraphics[width=\linewidth]{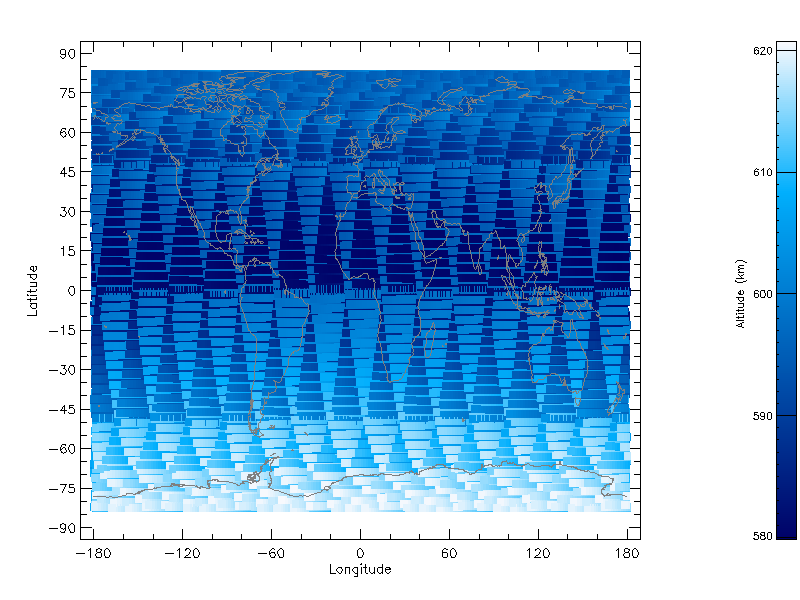}
    \caption*{Phase~1: SSO orbit.}
  \end{subfigure}
  \hfill
  \begin{subfigure}[b]{0.32\textwidth}
    \centering
    \includegraphics[width=\linewidth]{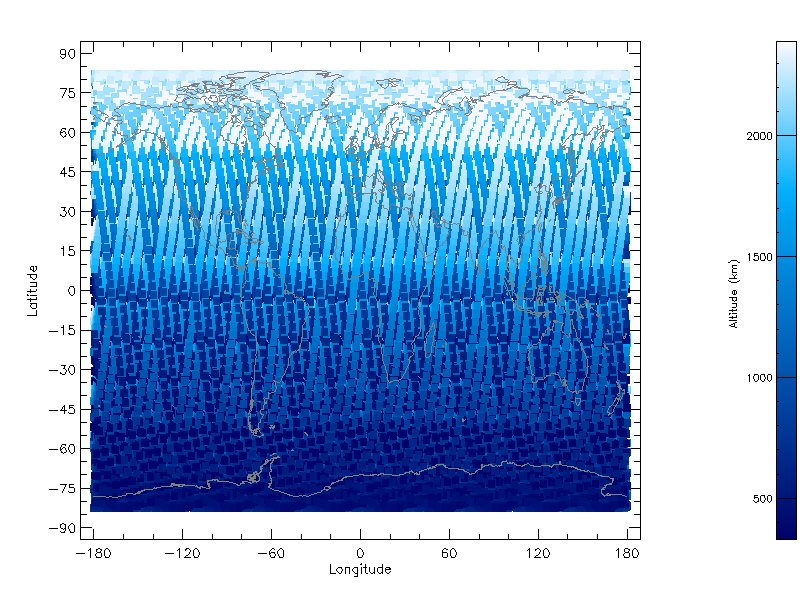}
    \caption*{Phase~2: Intermediate orbit.}
  \end{subfigure}
  \hfill
  \begin{subfigure}[b]{0.32\textwidth}
    \centering
    \includegraphics[width=\linewidth]{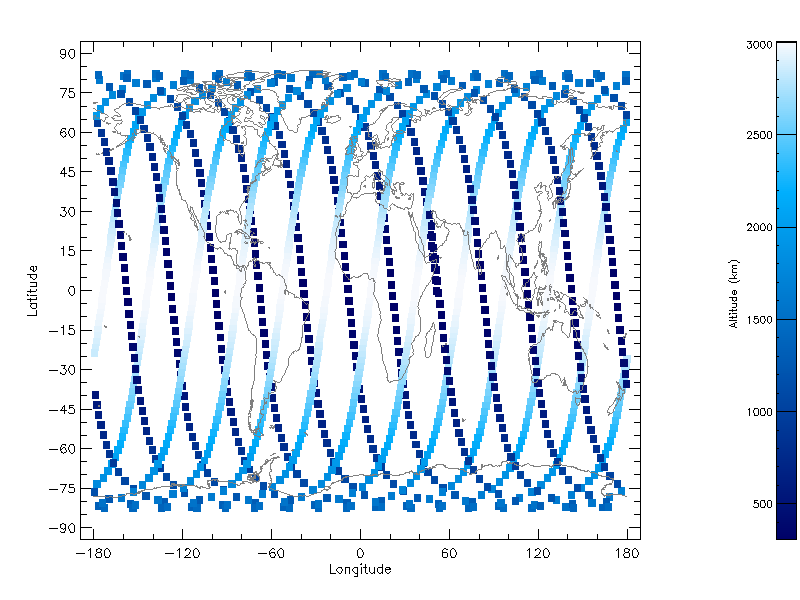}
    \caption*{Phase~3: Final orbit.}
  \end{subfigure}
  \caption{Altitude profile maps of the three ROMEO mission phases,
    derived from SPENVIS orbital simulations using the ROMEO mission
    phase specifications.
    Phase~1 is a \SI{600}{km} SSO; Phase~2 is a transfer ellipse
    ($2400\times337$~km); Phase~3 is the final inner-belt orbit
    ($3026\times314$~km).}
  \label{fig:orbits}
\end{figure}

RUBIK pursues two complementary science objectives that are both uniquely
enabled by the ROMEO mission profile.
The first objective is the spatial mapping of charged-particle rates along
the walking orbit.
As the ROMEO satellite traverses altitudes from \SI{600}{km} through the
inner radiation belt to above \SI{3000}{km} across its three mission
phases (Figure~\ref{fig:orbits}), RUBIK will record the hit rate and
track-direction distribution of charged particles as a continuous function 
of orbital altitude and position. 
Beyond the rate map itself, the reconstructed track directions allow a search for directional anisotropies in the high-energy particle flux, in 
particular of the GCR component, whose multi-GeV primaries arrive 
ballistically on the local scale and thus retain their true incident 
direction; accumulated over many events, this can reveal preferred 
directions or regions in near-Earth space with enhanced high-energy flux.
The lower detection threshold of approximately \SI{25}{\MeV} for protons is derived from simulations (see section \ref{sec:simulation}) of the current detector design structure and represents a first conservative estimation.

\begin{figure}[h]
  \centering
  \includegraphics[width=0.75\linewidth]{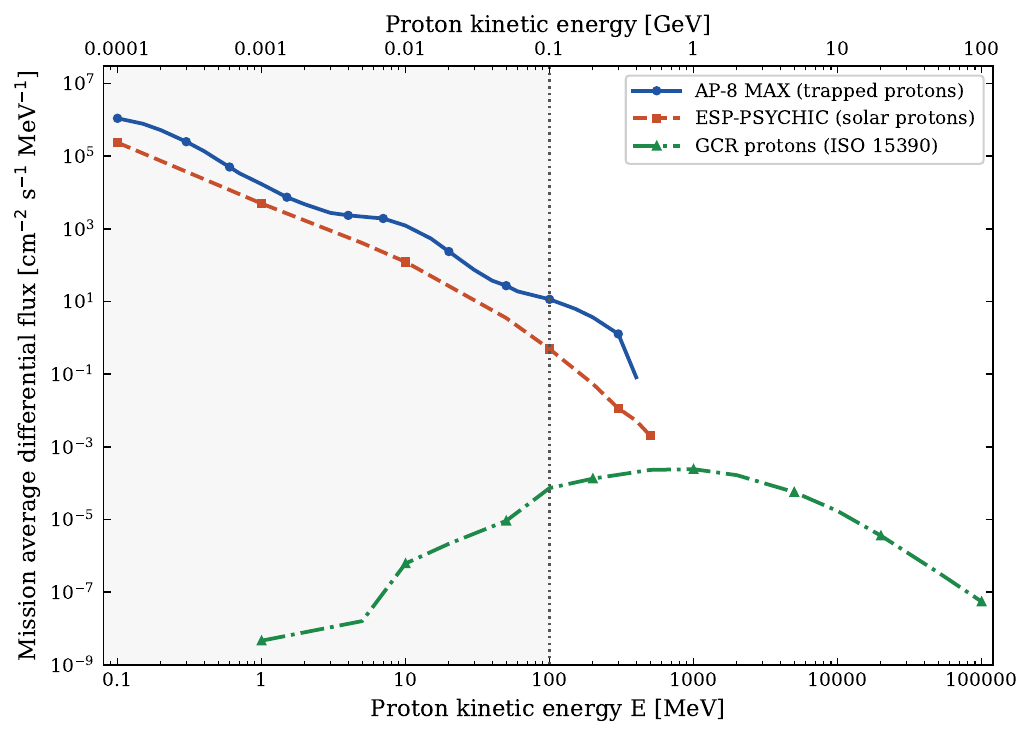}
  \caption{Mission-averaged differential proton flux (\SI{}{\per\square\centi\meter\per\second\per\MeV}) for the three particle populations: AP-8 MAX trapped protons, ESP-PSYCHIC solar protons (\SI{95}{\percent} confidence level, solar maximum), and ISO~15390 galactic cosmic rays. The shaded region and vertical dashed line indicate the effective detection threshold range of approximately up to \SI{100}{\MeV}, set by the five-fold coincidence 
requirement combined with the per-cube ToT criterion (Section~\ref{sec:simulation}).}
  \label{fig:flux_spectra}
\end{figure}

SPENVIS~\cite{heynderickx2004new,kruglanski2009space,kruglanski2009last_spenvis}
simulations using the AP-8 MAX trapped-proton
model~\cite{ap8,ap8ae8,ginet2013ae9}, the ESP-PSYCHIC solar-particle
model~\cite{xapsos1999probability,xapsos2000probability,xapsos2007model},
and the ISO~15390 galactic cosmic ray (GCR)
model~\cite{iso2004space} characterise three distinct proton populations
that determine the scientific environment.
The mission-averaged differential flux spectra for all three components,
evaluated over a 90-day simulation period, with 30 days allocated per mission phase to capture characteristic orbit-averaged flux profiles within the SPENVIS modeling framework, are shown in
Figure~\ref{fig:flux_spectra}.

\begin{figure}[h]
  \centering
  \includegraphics[width=0.75\linewidth]{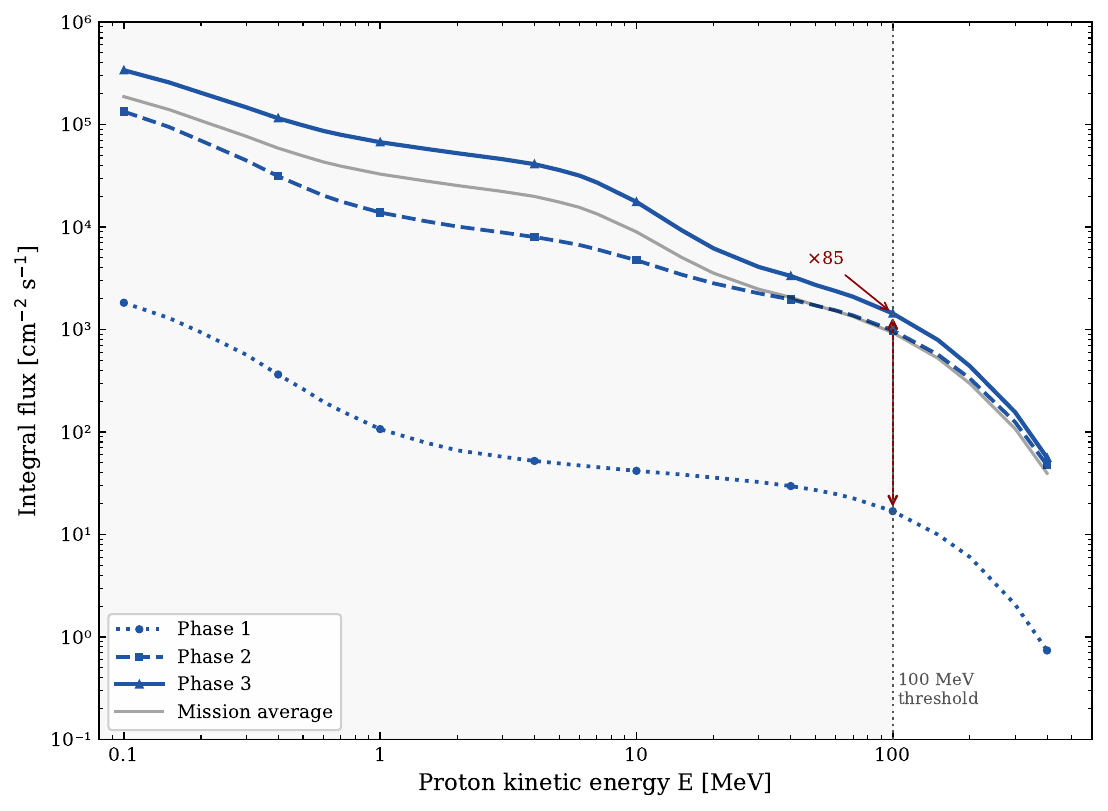}
  \caption{AP-8 MAX trapped proton integral flux for each of the three ROMEO mission phases, together with the mission-averaged spectrum (grey).}
  \label{fig:flux_spectra_per_phase}
\end{figure}

\textbf{Trapped protons (AP-8 MAX):}
The orbit-averaged integral proton flux above \SI{100}{MeV} increases from approximately \SI{17}{\per\square\centi\meter\per\second} in the initial \SI{600}{km} SSO (Phase~1) to \SI{982}{\per\square\centi\meter\per\second} in the intermediate orbit (Phase~2) and reaches \SI{1436}{\per\square\centi\meter\per\second} in the final orbit (Phase~3) --- a factor of approximately 85 between the lowest and highest altitude phases, with a mission average of \SI{935}{\per\square\centi\meter\per\second}. The phase-resolved differential flux spectra are shown in Figure~\ref{fig:flux_spectra_per_phase} and clearly illustrate the pronounced increase in trapped proton flux as the satellite enters the inner radiation belt during Phase~2 and Phase~3.

\textbf{Solar energetic particles (ESP-PSYCHIC):}
Based on ESP-PSYCHIC solar proton model simulations in SPENVIS over a one-year mission duration at \SI{95}{\percent} confidence level and solar maximum, the accumulated solar proton fluence above \SI{100}{\MeV}~--- the energy threshold above which protons traverse the full detector stack and constitute the primary GCR science target~--- is \SI{1.75e8}{\per\square\centi\meter}, corresponding to a time-averaged equivalent flux of approximately \SI{5.6}{\per\square\centi\meter\per\second}, approximately two orders of magnitude below the trapped proton component above the same threshold.

\textbf{Galactic cosmic rays (ISO~15390):}
The mission-averaged GCR proton flux above \SI{100}{\MeV\per\atomicmassunit} is \SI{1}{\per\square\centi\meter\per\second} (omnidirectional), increasing by approximately \SI{24}{\percent} from LEO to the final MEO orbit due to reduced geomagnetic shielding at higher altitudes. Although the GCR flux is dominated by the trapped proton environment by nearly three orders of magnitude in the mission average, GCR primaries --- protons, helium, and heavier nuclei above \SI{100}{\mega\electronvolt\per\atomicmassunit} constitute the scientifically relevant signal for the ground-to-space correlation objective of RUBIK, as they are the particles whose secondary cosmic-ray air showers are detectable at ground level.
This three-component hierarchy --- trapped protons $\gg$ solar protons
$\gg$ GCR above \SI{100}{\MeV} --- directly motivates the detector design:
below approximately \SIrange{80}{100}{\MeV}, the detector would be
overwhelmed by the trapped proton background; above this threshold the
strongly altitude-dependent trapped proton flux provides the rate gradient
that makes the walking-orbit profile of ROMEO scientifically unique.
The result is a spatially resolved rate map not accessible from any
fixed-altitude platform.
The second objective is to demonstrate the feasibility of ground-to-space
cosmic-ray correlation.
High-energy GCR primaries detected in orbit by RUBIK --- their arrival
direction reconstructed from the 3D hit pattern and their detection time
recorded relative to the OBC GPS time synchronisation --- may have
triggered an extensive cosmic-ray air shower observable at ground level at the same
instant.
RUBIK will provide the orbital side of this potential correlation.
To our knowledge, an event-by-event correlation of this kind has not
previously been attempted and would constitute the first experimental
proof of concept.

\section{Shielding and Energy Threshold Analysis}
\label{sec:simulation}

\begin{figure}[h]
  \centering
  \includegraphics[width=0.75\linewidth]{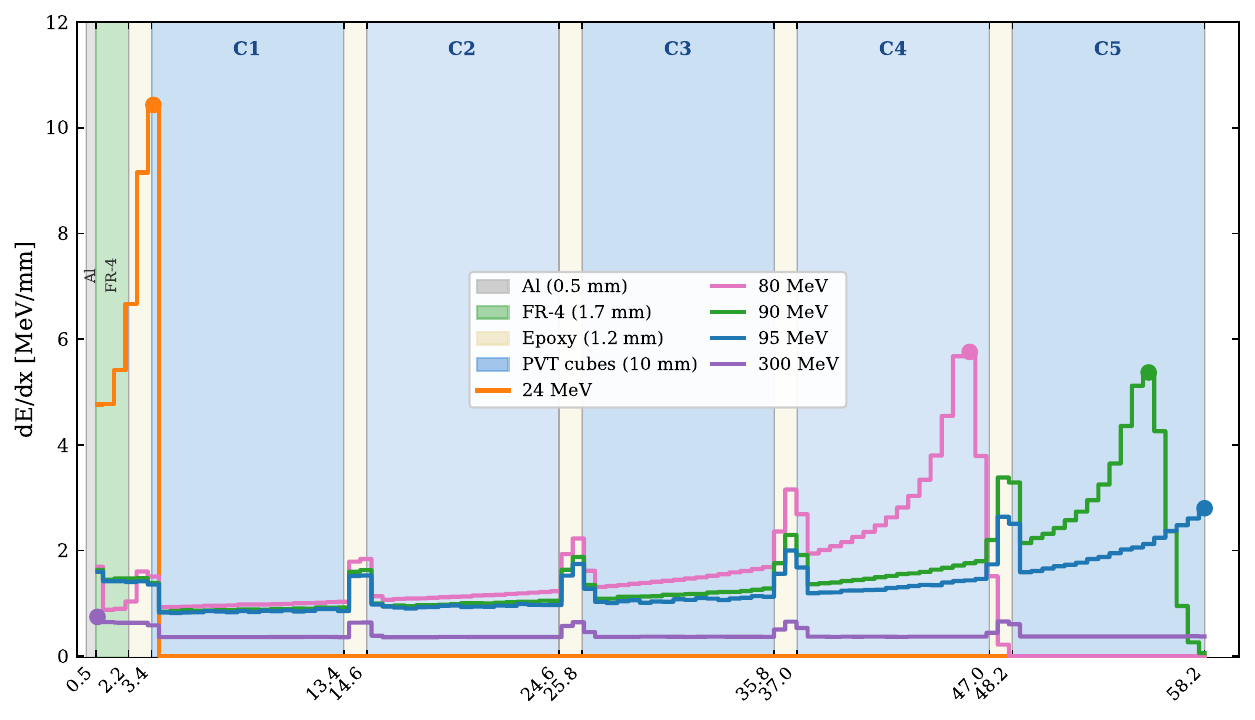}
  \caption{SRIM-2013 stopping-power profiles for representative proton energies incident perpendicular to the detector stack. Filled circles indicate the Bragg peak position; C1--C5 indicate the five successive PVT scintillator cubes along the particle path. Structural shielding of the satellite is not included.}
  \label{fig:srim}
\end{figure}

The detector response to the space radiation environment was characterised
by simulation approaches with the SRIM-2013~\cite{Ziegler2010} range--ionisation
calculation framework.
To determine the proton energy ranges accessible to the full detector
stack, ionisation profiles were calculated
for the complete one-dimensional geometry using SRIM-2013: \SI{0.5}{mm} Al,
\SI{1.7}{mm} FR4, and five alternating layers of \SI{1.2}{mm} epoxy and
\SI{10}{mm} PVT scintillator (cubes C1--C5), giving a total sensitive
depth of \SI{58.2}{mm}, consistent with the mechanical design described in Section~\ref{sec:detector}.
The resulting stopping power (dE/dx) profiles are shown in Figure~\ref{fig:srim}.

The results establish three physically motivated operating regimes:
\begin{itemize}
  \item Protons with kinetic energies below \SI{25}{\MeV} are fully stopped within the aluminum housing and FR4 layer and do not produce any signal in the sensitive volume: SRIM simulations predict Bragg peaks at \SI{2.3}{mm} (inside the Al shielding) and \SI{3.5}{mm} (at the FR4 / epoxy boundary) for \SI{20}{\MeV} and \SI{24}{\MeV} protons, respectively.

  \item Protons with kinetic energies between \SI{25}{\MeV} and $\approx$ \SI{80}{\MeV} enter the sensitive volume but stop before reaching the most distal layer C5 (at normal incidence). At \SI{80}{\MeV} the Bragg peak is located at a penetration depth of \SI{45.4}{mm}, within C4, so C5 does not receive a signal; these events provide a unique signature to identify this energy regime.

  \item Protons with kinetic energies between approximately \SI{80}{\MeV} and \SI{100}{\MeV} impinge on all five cube layers at normal incidence but stop within C5, depositing their full remaining energy there. Although these events satisfy the five-fold coincidence, they are distinguishable from fully-traversing minimum-ionising particles by their ToT signature: a stopped proton produces a pronounced Bragg-peak enhancement in the last traversed cube, resulting in a larger ToT value in C5 relative to C1--C4.    The combination of the five-fold coincidence topology and individual ToT readings therefore extends the effective particle discrimination up to approximately \SI{100}{\MeV}.

  \item Protons with kinetic energies above \SI{100}{\MeV}, such as GCRs, traverse the entire detector stack without being stopped. The hardware discriminator threshold must be set low enough to detect the minimum ionising signal, corresponding to approximately \SI{2}{\MeV} deposited energy per cube. Primary background rejection relies on the requirement of a coincident signal in all five layers (multiplicity-5 trigger) combined with the ToT uniformity criterion.
\end{itemize}

\section{The RUBIK Detector}
\label{sec:detector}

\begin{figure}[h!]
  \centering
  \includegraphics[width=0.65\linewidth]{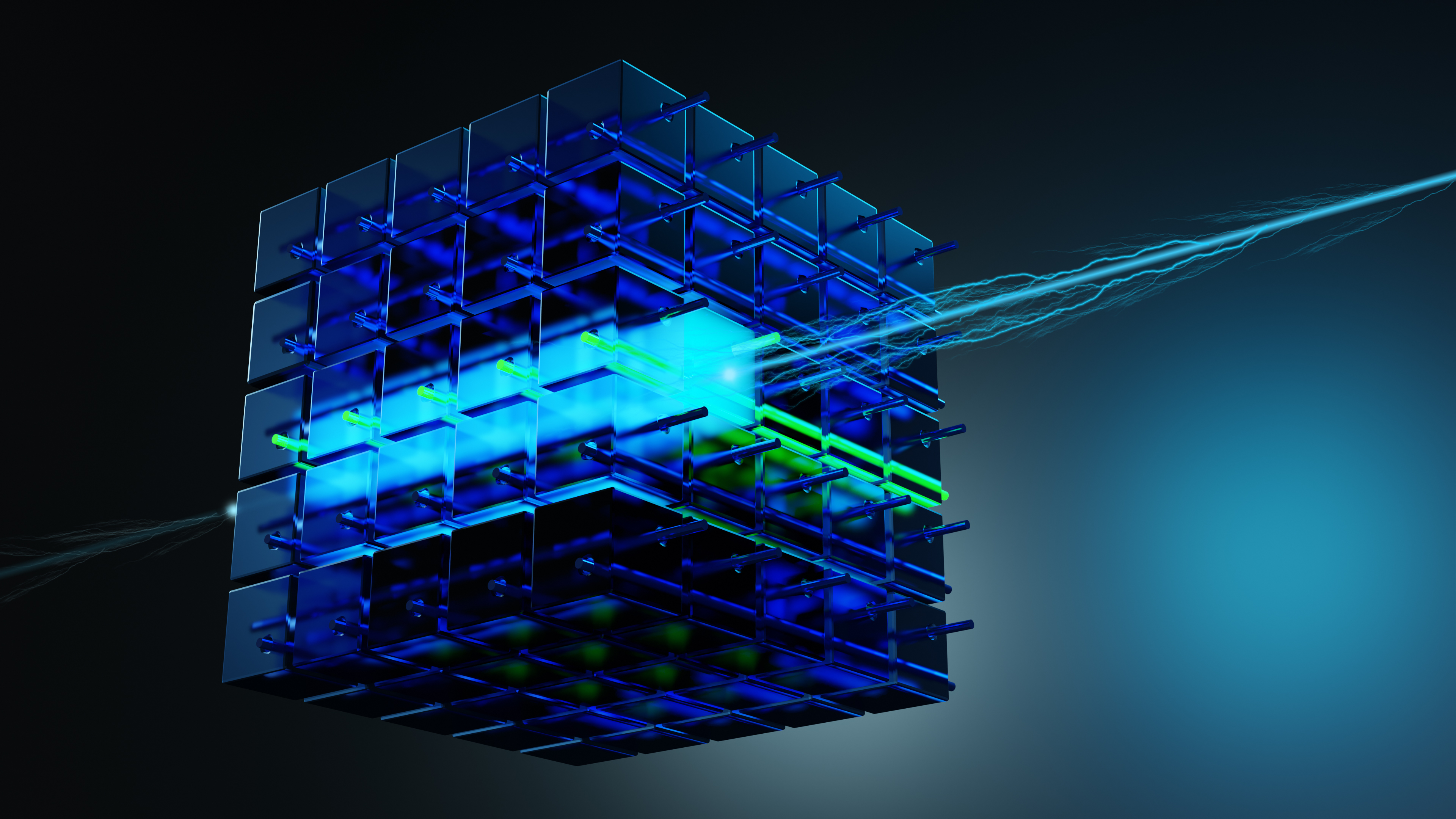}
  \caption{Graphical representation of the 125-cube sensitive volume
    penetrated by WLS fibres, illustrating the track-reconstruction logic.
    Credit: Elisa Monte/JLU Giessen.}
  \label{fig:tracking}
\end{figure}

RUBIK consists of a \(5{\times}5{\times}5\) array of plastic PVT-based scintillator
unit cubes read out via wavelength-shifting (WLS) fibres (Figure \ref{fig:tracking}) and silicon
photomultipliers (SiPMs), housed in a sealed aluminum frame.
The readout electronics comprise three custom PCBs: two identical
discriminator and SiPM preamplifier boards forming the analogue front-end,
one FPGA-based data-acquisition and communication board, and one power and
bias distribution board.
A single external connector links the detector to the satellite bus; an
RS-422 differential link carries science and housekeeping data, a 1\,PPS
signal from the on-board computer (OBC) provides clock synchronisation,
and the satellite power system supplies nominal \SI{28}{V}.

\subsection{Segmented Sensitive Volume}
\label{subsec:volume}

\begin{figure}[h!]
  \begin{subfigure}[b]{0.55\textwidth}
    \centering
    \includegraphics[width=\linewidth]{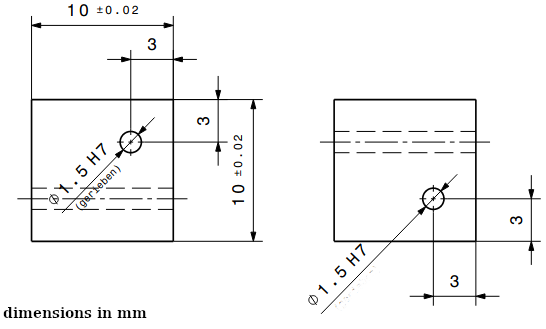}
    \caption{Unit cube design showing the two perpendicularly drilled fibre
      channels.}
  \end{subfigure}
  \hfill
  \begin{subfigure}[b]{0.42\textwidth}
    \centering
    \includegraphics[width=\linewidth]{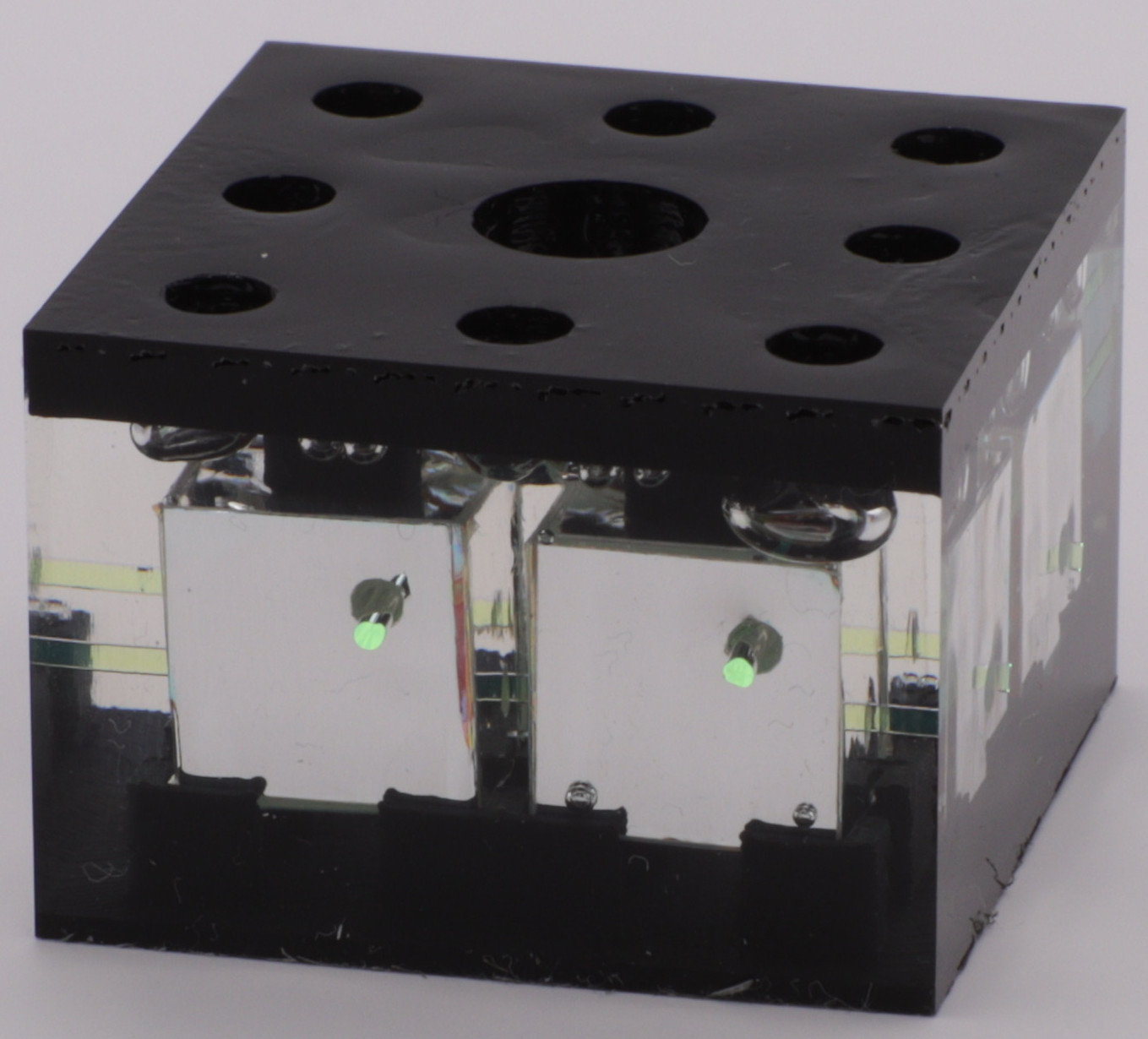}
    \caption{Cast and machined prototype block showing the epoxy sealing
      and flat-machined fibre end-faces.}
  \end{subfigure}
  \caption{Unit cube and first cast prototype of the RUBIK sensitive
    volume.}
  \label{fig:cube}
\end{figure}

The sensitive volume is based on 125 unit cubes of EJ-248 (Eljen
Technology), a PVT-based plastic scintillator formulated for elevated
thermal endurance (softening point $\approx$\SI{99}{\celsius}) with a
peak emission wavelength of \SI{425}{nm}~\cite{ref:ej-248}, well-matched
to the WLS absorption maximum of \SI{430}{nm}.
Each cube has a side length of \SI{10}{mm} and two perpendicularly aligned
fibre channels drilled through adjacent corners (Figure~\ref{fig:cube}).
The granular configuration is inspired by the fully active
scintillator concept of Blondel et
al.~\cite{blondel2018fully,mineev2019beam,blondel2020superfgd}, adapted
here to a two-projection readout geometry and a simplified cube wrapping
scheme.
The outer faces of each cube are wrapped in DF-2000MA (3M), a specular
reflective film~\cite{ref:df-2000}, to maximise photon collection
efficiency and suppress optical crosstalk between adjacent cubes.

The cubes are penetrated by \SI{1}{mm} Kuraray Y-11 wavelength-shifting
fibres (polystyrene core, PMMA cladding)~\cite{ref:wls} running along two
orthogonal axes --- 25 fibres per projection ($x$ and $y$).
Each fibre absorbs the blue scintillation photons emitted by the PVT
matrix and re-emits them at green wavelengths, guiding the light to a SiPM photosensor at the fibre end.
The two-projection readout with 25 channels per axis results in 50 independent channels in total and enables reconstruction of the full 3D hit pattern from coincident activation of fibres in both projections.

The entire assembly --- wrapped cubes and fibres --- is cast into a
monolithic block using EPO-TEK\textsuperscript{\textregistered}~301-2
(Epoxy Technology) optical-grade transparent epoxy resin (NASA outgassing
compliant), cured at \SI{80}{\celsius} for \SI{3}{h}~\cite{ref:epoxy}.
The casting case is subsequently machined away to the final detector
dimensions, yielding flat end-faces for direct SiPM attachment to the
fibre ends.

\subsection{Mechanical Design}
\label{subsec:mechanical}

The detector is enclosed in an aluminum frame (EN AW-7019 alloy,
\SI{0.5}{mm} sheet panels) that simultaneously serves as ambient light protection, partial radiation shielding, and thermal interface to the satellite
structure.
The current mass estimate is \SI{914}{g} for the full assembly including
all electronics and housing components; including the assembly tolerances the final mass is expected to remain below \SI{1}{kg}.
As nearly half the mass is concentrated in the sensitive volume, the
centre of mass lies close to the geometric centre of the cubic envelope,
with a small offset along the $z$-axis.

\begin{figure}[h!]
  \begin{subfigure}[b]{0.45\textwidth}
    \centering
    \includegraphics[width=\linewidth]{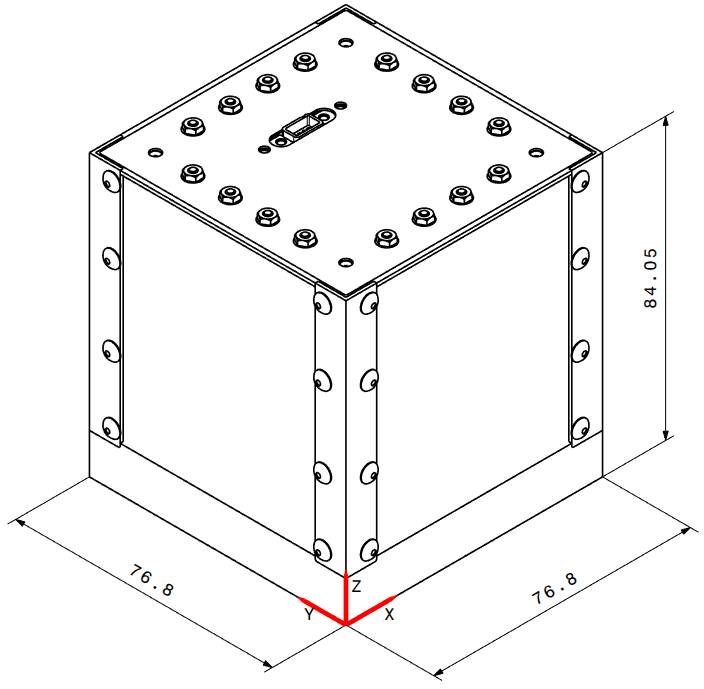}
    \caption{Dimensional overview of the aluminum-enclosed design. (Dimensions are in mm)}
  \end{subfigure}
  \hfill
  \begin{subfigure}[b]{0.55\textwidth}
    \centering
    \includegraphics[width=\linewidth]{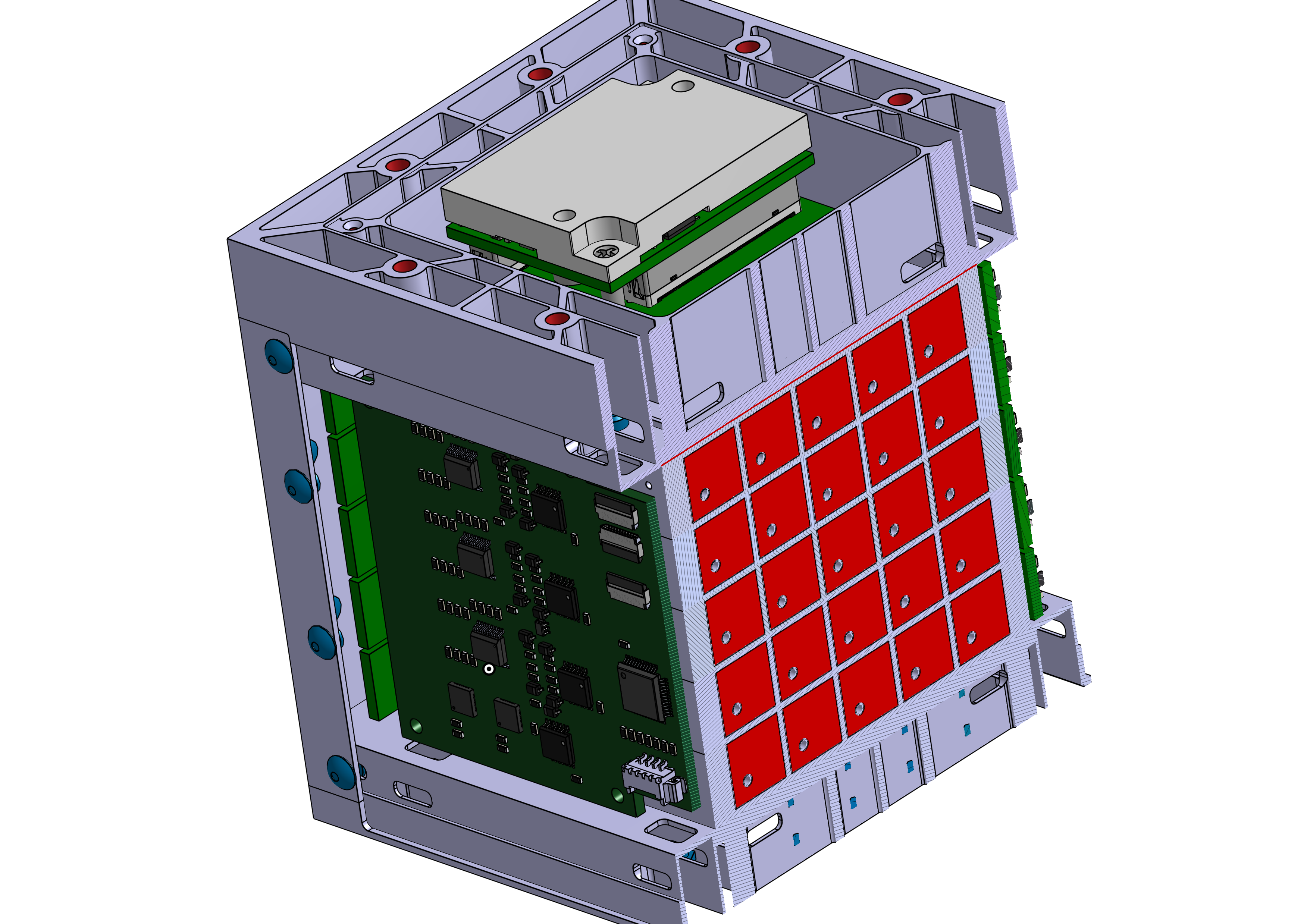}
    \caption{Cross-sectional view showing the FPGA module with heat
      spreader at the top.}
  \end{subfigure}
  \caption{Mechanical design of the RUBIK detector enclosure. (a) The aluminum 
enclosure with the satellite interface connector opening visible on 
the top face. (b) computer-aided design (CAD) overview of the full assembly.}
  \label{fig:mechanical}
\end{figure}

Thermal management is passive: each PCB is thermally coupled to the metal
frame by a copper heat sink, and a heat spreader is mounted on the FPGA
module (see the top face in Figure \ref{fig:mechanical} (b)).
The detector attaches to the satellite structure via four M4 screw threads
on the outer housing faces; the detector face at the fastening interface
also serves as the thermal conduction path towards the satellite structure.

\subsection{Signal Readout Chain}
\label{subsec:readout}

\begin{figure}[h]
  \centering
  \includegraphics[width=0.95\linewidth]{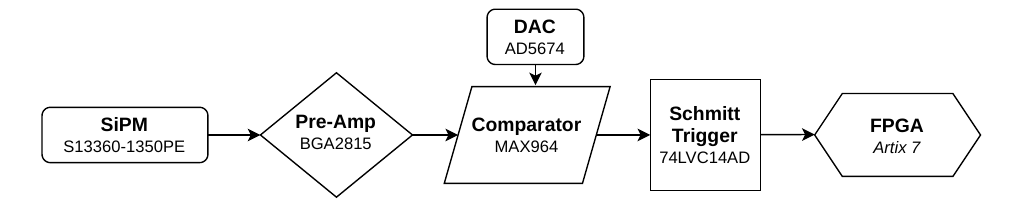}
  \caption{Single-channel signal readout chain from the SiPM photosensor until the digital capture by the FPGA unit.}
  \label{fig:readout_chain}
\end{figure}

Each WLS fibre is read out by a $1.3 \times 1.3~\mathrm{mm^2}$ Hamamatsu
S13360-1350PE SiPM \cite{ref:sipm} mounted on a custom PCB that positions the SiPMs
precisely at the fibre end-faces.
A wideband RF preamplifier (Section~\ref{subsec:discr}) is placed on the
opposite face of the PCB, with a ground-via fence per channel for isolation.
This SiPM--preamplifier assembly connects via a flex-print PCB
(FR4/polyimide substrate) to the discriminator board, where
leading-edge comparators digitise the analogue signals. The readout chain for a single channel, from the analogue SiPM signal to the 
digital signal processed by the FPGA, is illustrated in Figure~\ref{fig:readout_chain}. The nominal SiPM operating voltage is \SI{56}{V}, supplied
and monitored by the power and bias distribution board
(Section~\ref{subsec:power}).
The quiescent total SiPM bias current does not exceed
\SI{10}{\micro A} under nominal conditions.
A radiation-induced increase in SiPM dark current is anticipated for the radiation belt environment and has been quantified on the prototype via total ionising dose (TID) measurements up to an accumulated dose of \SI{150}{\gray}. The observed degradation, primarily an increase in dark current and hence dark count rate, is consistent with effects reported in the literature~\cite{garutti2019radiation,zheng2022orbit,burmistrov2025performance} for SiPMs under comparable irradiation conditions.

\section{Electronics Design}
\label{sec:electronics}

Three dedicated electronic subsystems provide analogue pulse processing, digitisation, event formatting, monitoring and control. These are all based exclusively on COTS components, in line with the design philosophy of the mission.
The two discriminator boards provide the analogue front-end for 25
channels each.
The FPGA board handles data acquisition, event formatting, and
communication with the satellite OBC.
The power and bias board generates all supply and SiPM bias voltages and
monitors system status.
The boards interconnect via Hirose flat flexible cable (FFC) connectors and FFC cables to the
FPGA board. Power is internally distributed via Molex PicoBlade connectors and cable assemblies.

\subsection{Power and Bias Distribution}
\label{subsec:power}

The detector receives \SI{28}{V} (nominal satellite bus supply voltage) via the
external Harwin G125-2241096F5 \cite{ref:gecko} (gecko family) ten-pin connector (Table~\ref{tab:connector}).
Three LMR33620 synchronous buck converters \cite{ref:lmr33620} step the input down to
\SI{3.3}{V} for the electronics boards.
The nominal \SI{56}{V} SiPM bias voltage is generated by an LT8361 boost converter \cite{ref:lt8361} from the \SI{28}{V} input.
Fine adjustment of the bias voltage is achieved via a RT9072B LDO \cite{ref:ldo} regulated by the DAC output of the ATtiny816 \cite{ref:attiny}.
The bias supply topology is derived from the open-source
\textit{OpenSiPM} design~\cite{ref:biaspcb_github,ching2021ultrahigh},
adapted to the required \SI{56}{V} output and the Microchip ATtiny816
control loop.

The Microchip ATtiny816 microcontroller serves as power
sequencer and housekeeping monitor.
Operating as an serial peripheral interface (SPI) slave to the FPGA master, it executes a four-stage
power-enable finite state machine with \SI{2}{s} per-stage timeouts and
power-good validation.
Housekeeping telemetry --- bias voltage (averaged ADC reading of a passive voltage divider),
temperature from the internal sensor, power-good status, and set bias
voltage --- is returned to the FPGA on request.

\begin{table}[htb]
  \centering
  \caption{Pin assignment of the external interface connector
    (Harwin G125-2241096F5, 10-pin) between RUBIK and the satellite bus.}
  \label{tab:connector}
  \begin{tabular}{cll}
    \toprule
    \textbf{Pin} & \textbf{Signal} & \textbf{Description} \\
    \midrule
    1 & Rx\_P  & RS-422 receive (positive) \\
    2 & Rx\_N  & RS-422 receive (negative) \\
    3 & Tx\_P  & RS-422 transmit (positive) \\
    4 & Tx\_N  & RS-422 transmit (negative) \\
    5 & PPS\_P & 1\,PPS time reference (positive) \\
    6 & PPS\_N & 1\,PPS time reference (negative) \\
    7 & NC     & Not connected \\
    8 & GND    & Power/signal return \\
    9 & +28\,V & Nominal satellite bus supply \\
    10& GND    & Power/signal return \\
    \bottomrule
  \end{tabular}
\end{table}

\subsection{Data Acquisition Board}
\label{subsec:fpga}

The data acquisition board is built around the TE0714 module (Trenz Electronic)
carrying a Xilinx Artix-7 XC7A50T FPGA~\cite{ref:fpga}.
The module integrates a \SI{16}{MByte} QSPI boot flash device loaded
with two bitstreams stored at separate flash addresses: a golden
fallback image and a primary operational bitstream.
Xilinx MultiBoot autonomous reconfiguration is implemented to recover for example from
bitstream corruption caused by single-event upsets (SEUs) in the
radiation belt environment.

The FPGA is the sole SPI master on a shared bus serving seven slaves:
four AD5674R DACs \cite{ref:dac} (two per discriminator board, providing a total of
50 threshold channels), two LTC2986 temperature sensors (one per
discriminator board), and the ATtiny816 microcontroller on the power
board.

\subsection{Front-end Electronics}
\label{subsec:discr}

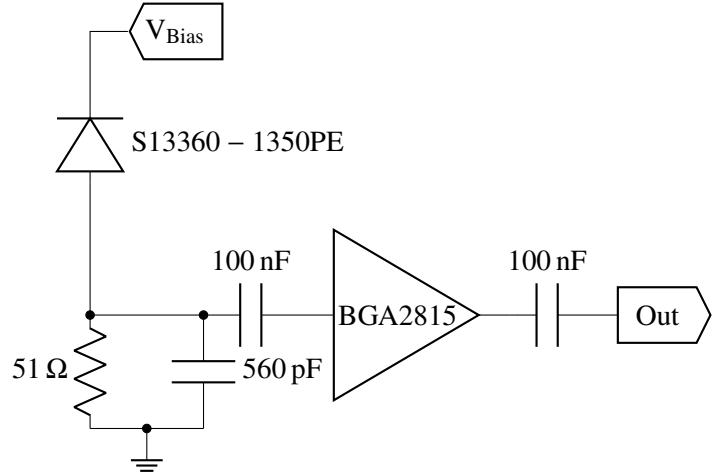
\begin{figure}[h]
\centering
\input {preamp_rubik.tex}
\caption{Schematics of RF amplifier and SiPM frontend.}
\label{fig:preamp}
\end{figure}

Each front-end board services 25 SiPM channels.
The preamplifier stage (Figure \ref{fig:preamp}) uses a wideband RF MMIC (monolithic microwave
integrated circuit) from NXP's BGA28XX series~\cite{ref:bga2815}, which is internally matched to \SI{50}{\ohm}. 
Two devices from this family were evaluated during prototyping: the
BGA2800~\cite{ref:bga2800} (current consumption $\approx$\SI{10}{mA},
output 1\,dB compression point $P_\mathrm{1dB} \approx -1$\,dBm) and the
BGA2815~\cite{ref:bga2815} ($\approx$\SI{16}{mA},
$P_\mathrm{1dB} \approx +5$\,dBm).
The BGA2815 was selected for the final design due to its higher output voltage swing, which accommodates the large pulse amplitudes expected from stopped-particle interactions in the sensitive volume. Similar MMIC-based preamplifier front-ends have been realised in house for a proton range telescope \cite{fiedler2026pbwo4} and for a portable, compact neutron detector \cite{kazlou2026red}. 

An RC shaping network consisting of the \SI{51}{\ohm} sense resistor in parallel to a \SI{560}{pF} capacitance is placed at the preamplifier input. This network integrates the fast SiPM current pulse, which typically exhibits a rise time of a few nanoseconds and a decay dominated by individual Geiger-mode avalanche contributions, into a smoothed voltage pulse with an average pulse width of approximately \SI{30}{ns}.
The integration suppresses the discrete single-photoelectron peak structure characteristic of SiPM signals, which would otherwise cause multiple threshold crossings per event with potential degradation of the time-over-threshold measurement.
Therefore, the resulting shaped pulse provides a well-defined, single threshold crossing on both the leading and falling edges, enabling a stable and reproducible ToT measurement proportional to the total integrated charge -- and hence to the deposited
energy in the scintillator.

Temperature monitoring on each discriminator board is provided by an
LTC2986 (Analog Devices) multi-sensor IC~\cite{ref:ltc2986} with TMP235
analogue temperature sensors (Texas Instruments)~\cite{ref:tmp235}.
The temperature readings are forwarded via SPI communication to the FPGA and serve as input
to two complementary compensation strategies: the SiPM bias voltage can
be adjusted via the power board control loop to maintain a stable
operating overvoltage across varying temperatures, compensating for the
temperature dependence of the SiPM breakdown voltage; alternatively, the
per-channel discriminator thresholds can be adjusted via the DAC to
account for the corresponding gain shift.
Both strategies can be applied independently or in combination depending
on the operating scenario.

\section{Data Acquisition and Communication}
\label{sec:daq}

The primary science data is a variable-length event record representing all channels that fires within a configurable coincidence time window, comprising a 48-bit event timestamp followed by $N_\mathrm{hit}$ channel hit records of 12\,bits each, where $N_\mathrm{hit}$ denotes the number of fired channels in the event. Each channel record 
encodes a 6-bit channel identifier (range 0--49) and a 6-bit 
time-over-threshold value, with one record per fired channel. 

The 48-bit timestamp is partitioned into a 20-bit coarse seconds counter 
and a 28-bit sub-second fine counter. This yields a deterministic timing
resolution of exactly $1 / 2^{28}$~seconds ($\approx 4 \text{ ns}$), 
matching the target sub-10~ns precision envelope. The hardware implementation 
uses an external GPS disciplined pulse per second (PPS) signal to latch the coarse counter boundary and synchronously reset the fine-grain LSB vector, effectively isolating orbital thermal oscillator drifts.

The total event size is governed by the byte-aligned ceiling function
\begin{equation}
    L_\mathrm{event} = \left\lceil \frac{48 + 12\,N_\mathrm{hit}}{8} 
    \right\rceil \,\mathrm{bytes},
\end{equation}
yielding 6\,bytes for $N_\mathrm{hit} = 0$ and an average of 15\,bytes 
for $N_\mathrm{hit} = 6$ representing a five unit cube coincidence hit pattern. At the expected nominal event rate of \SI{127}{Hz}, estimated from the SPENVIS simulation for protons, 
following a hardware coincidence filter and assuming a typical hit 
multiplicity of six channels per event, this corresponds to a data 
throughput of approximately \SI{1.9}{kB/s}.

The 6-bit ToT field encodes the duration for which each channel 
signal exceeds the discriminator threshold, sampled at 
\SI{250}{MHz} (4\,ns resolution per bin), providing a proxy for 
deposited energy per channel. The 6-bit word accommodates pulse 
durations up to \SI{252}{ns} (63 clock cycles).

Cube-level energy reconstruction is performed in ground-side analysis software by combining the ToT values of the X and Y WLS fibre channels 
whose identifiers satisfy the coincidence condition
\begin{equation}
    \mathrm{Cube}(i,\,j,\,L) \text{ hit} \iff 
    \mathrm{Ch}[10L+i] \in \mathcal{F} 
    \;\wedge\; 
    \mathrm{Ch}[10L+5+j] \in \mathcal{F},
\end{equation}
where $\mathcal{F}$ denotes the set of fired channel identifiers in the 
event record and $L \in \{0,\ldots,4\}$ denotes the detector layer.

The telemetry framework is planned to accommodate periodic housekeeping packets comprising time-averaged count rates, bias voltage, and subsystem temperatures, subject to internal block random access memory (BRAM) constraints. Diagnostic status flags are foreseen to verify DAC threshold configurations while minimising packet overhead.

Data acquisition and readout are coordinated entirely over SPI: the FPGA-based DAQ board communicates with the discriminator boards for threshold configuration, the bias supply for SiPM voltage control, and the onboard temperature sensors for housekeeping. External communication with the ROMEO satellite bus is implemented over a differential RS-422 serial link, following the Consultative Committee for Space Data Systems (CCSDS) packet standard \cite{ref:ccsds}, encapsulating both science event records and periodic housekeeping telemetry into CCSDS Space Packets for downlink.

\section{Prototype Studies and First Results}
\label{sec:proto}

\begin{figure}[h!]
  \begin{subfigure}[b]{0.45\textwidth}
    \centering
    \includegraphics[width=\linewidth]{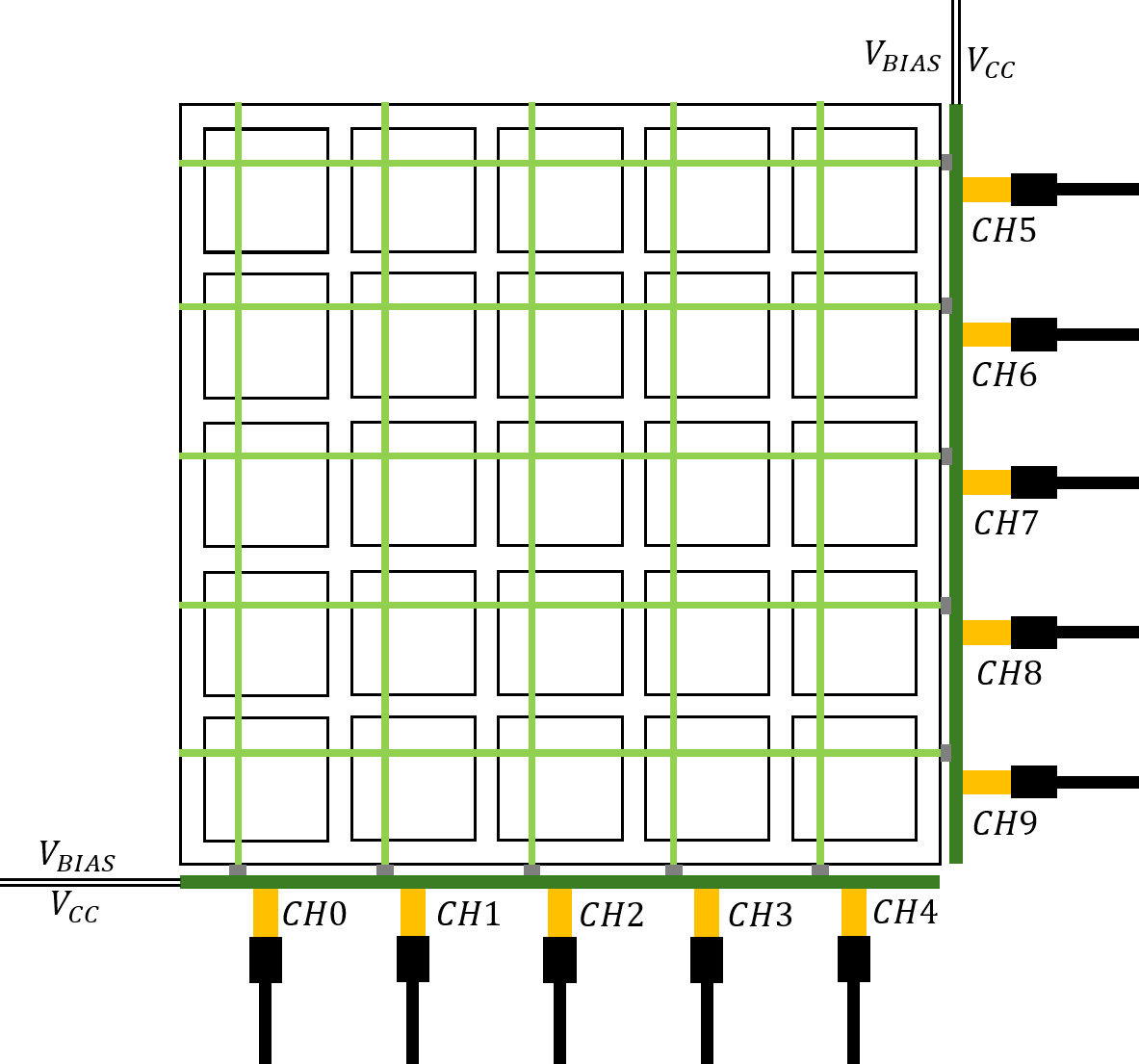}
    \caption{Schematic of the \(5{\times}5{\times}1\) prototype readout
      geometry with embedded WLS fibre structure.}
  \end{subfigure}
  \hfill
  \begin{subfigure}[b]{0.45\textwidth}
    \centering
    \includegraphics[width=\linewidth]{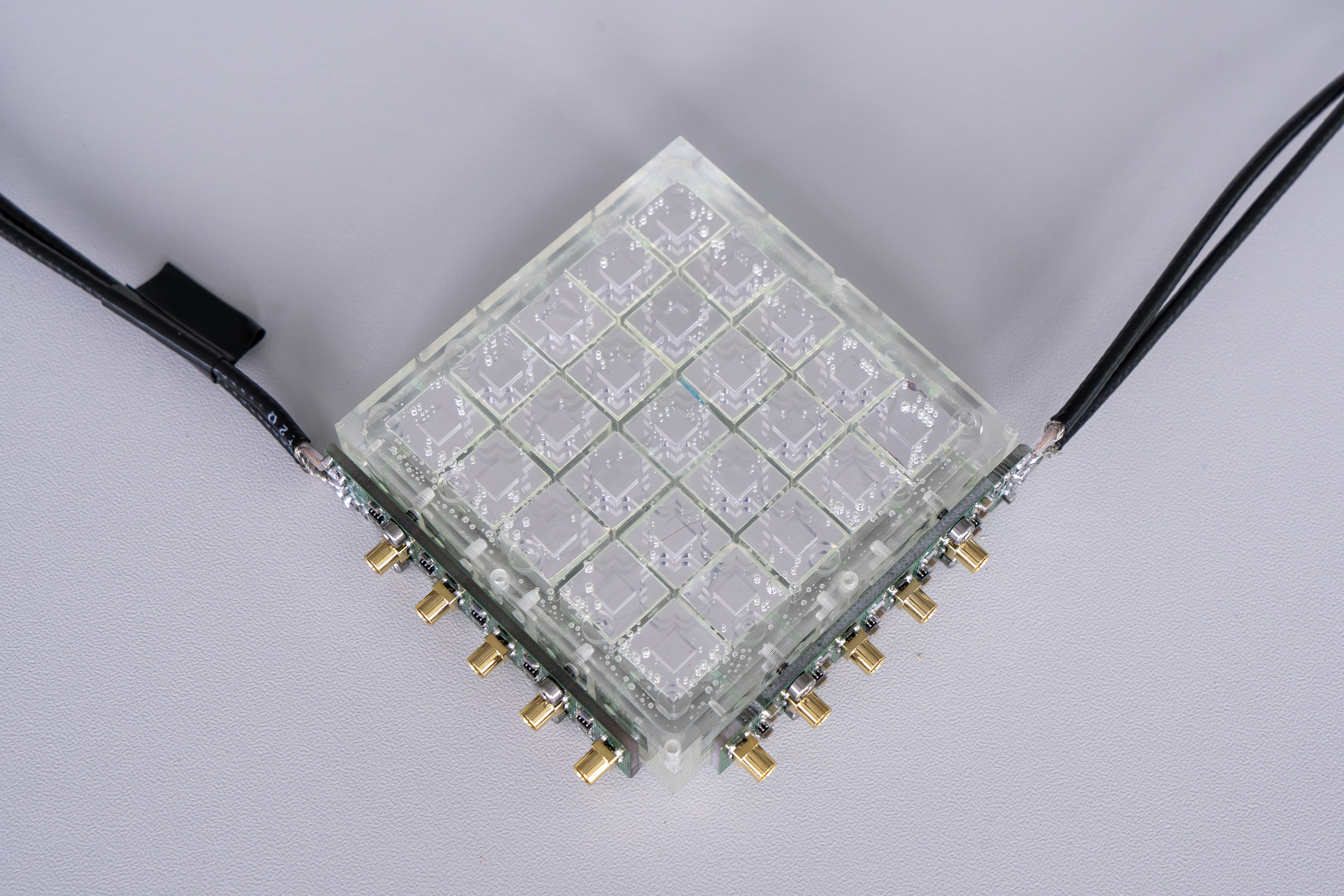}
    \caption{Photograph of the assembled \(5{\times}5{\times}1\) single-layer prototype.}
  \end{subfigure}
  \caption{The \(5{\times}5{\times}1\) single-layer prototype used for all
    characterisation measurements.}
  \label{fig:prototype}
\end{figure}

A \(5{\times}5{\times}1\) single-layer prototype (Figure \ref{fig:prototype}) was constructed to validate core design decisions prior to assembly of the full \(5{\times}5{\times}5\) array. The prototype is a geometrically identical slice of the flight model's sensitive volume: 25 EJ-248 scintillator cubes cast with Kuraray Y-11 fibres in
EPO-TEK~301-2 epoxy resin, read out via Hamamatsu S13360-1350PE SiPMs and BGA2815 RF amplifiers, providing ten readout channels (five per readout projection).

\subsection{Temperature and Bias Dependence}
\label{subsec:sipm_char}

The dependence of each channel on the operating temperature and bias voltage 
of the SiPMs was characterised using a $^{90}$Sr beta source and a Weiss 
WT11\,180/40 climate chamber. These measurements were recorded with a CAEN 
V1730S waveform digitiser (14-bit ADC, \SI{500}{MS/s}), which allowed all ten channels to be acquired simultaneously to characterise the general response of the full prototype readout.

\begin{figure}[h]
  \begin{subfigure}[b]{0.49\textwidth}
    \centering
    \includegraphics[width=\linewidth]{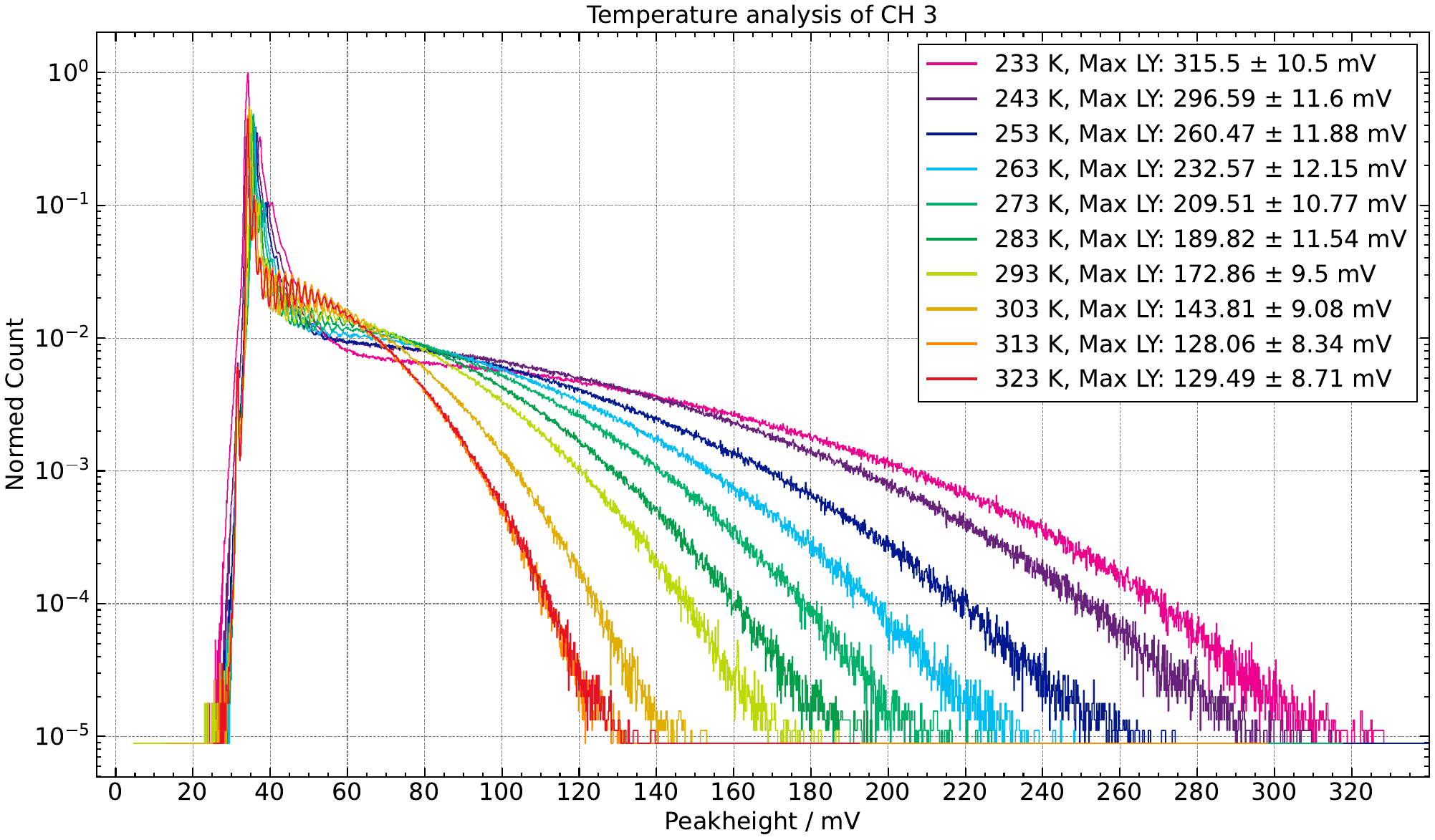}
    \caption{Temperature dependence of the amplitude spectrum at fixed SiPM bias voltage.}
  \end{subfigure}
  \hfill
  \begin{subfigure}[b]{0.49\textwidth}
    \centering
    \includegraphics[width=\linewidth]{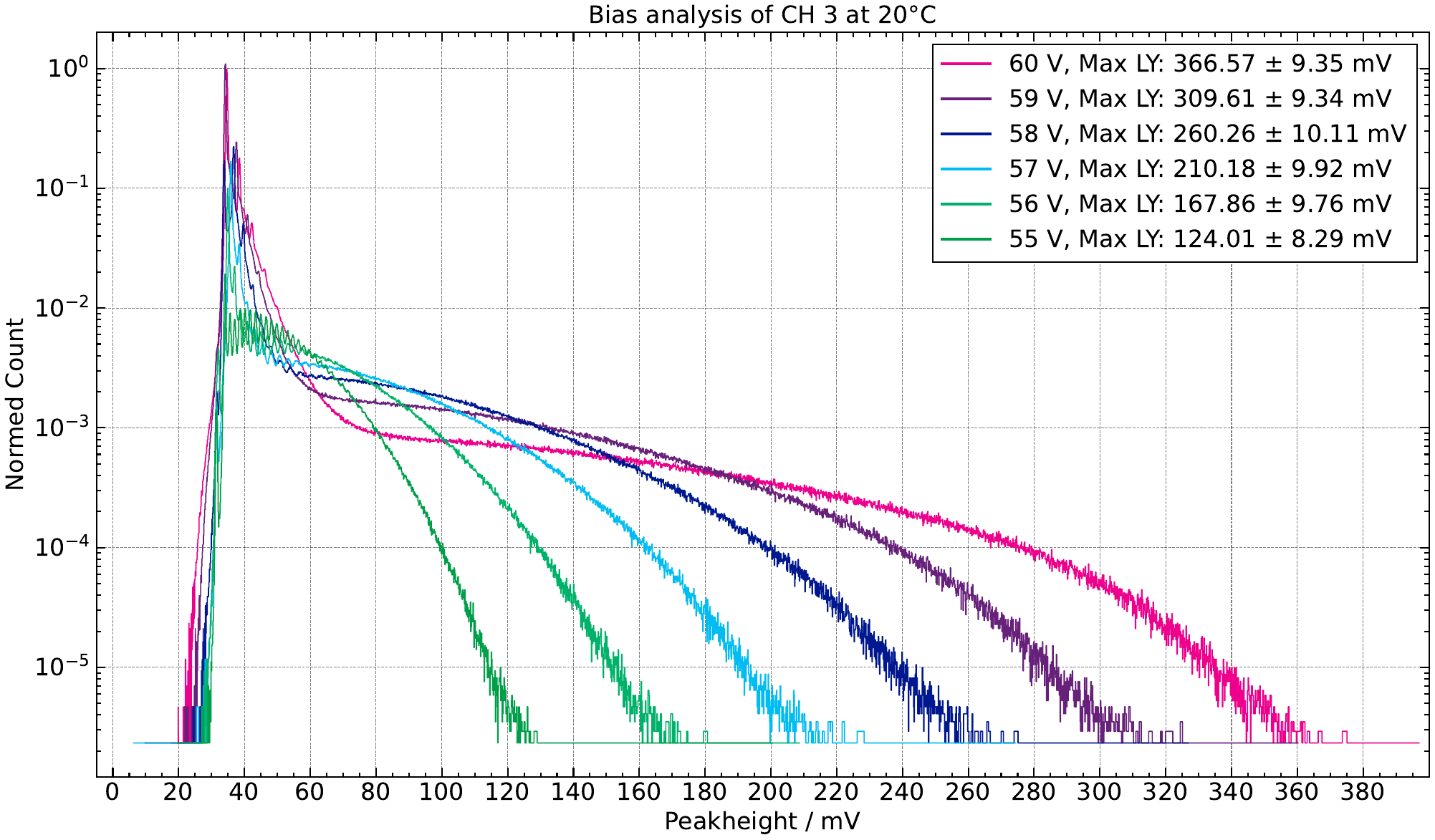}
    \caption{Amplitude spectrum for different bias voltage settings at \SI{20}{\celsius}.}
  \end{subfigure}
  \caption{SiPM pulse height distributions as a function of (a) temperature and (b) bias voltage measured with a $^{90}$Sr beta source (channel~3 shown as representative example).}
  \label{fig:sipm_char}
\end{figure}

Two clear trends were identified: at a fixed bias voltage, an increase in
temperature reduces the SiPM gain (the overvoltage decreases as the
breakdown voltage rises with temperature, with an expected temperature coefficient of \SI{54}{\milli\volt\per\celsius} \cite{ref:sipm}), and there is an additional degradation in light output of the EJ-248 unit cubes and WLS fibres, which leads to lower combined signal amplitudes. A similar procedure was performed at a fixed temperature and increased bias voltage, which raises the gain and therefore the detected signal amplitude.
Representative results for channel~3 are shown in Figure~\ref{fig:sipm_char}; the upper edge of the spectrum was used as the benchmark quantity for comparing across settings.
These dependencies will be accounted for during RUBIK's in-orbit operation via a regulation loop that continuously receives temperature feedback from each analogue front-end board. 

To resolve the individual photoelectron peaks, the spectra presented in Figure~\ref{fig:sipm_char} were acquired in a dedicated diagnostic configuration in which the SiPM and RF preamplifier operate without the shaping capacitance parallel to the input resistor of the preamplifier (Section~\ref{subsec:discr}). In the deployment configuration this capacitance is added (see Figure~\ref{fig:preamp}), deliberately stretching the SiPM pulse --- reducing its amplitude while extending its duration at constant charge --- to establish the stable pulse-width baseline required for linear time-over-threshold digitisation. Importantly, this shaping modifies only the pulse amplitude and length; the underlying photoelectron structure is preserved and remains resolvable, as demonstrated by the calibration of the deployment configuration in the following subsection.

\subsection{Signal Calibration}
\label{subsec:calib}

\begin{figure}[h]
  \centering
  \includegraphics[width=\linewidth]{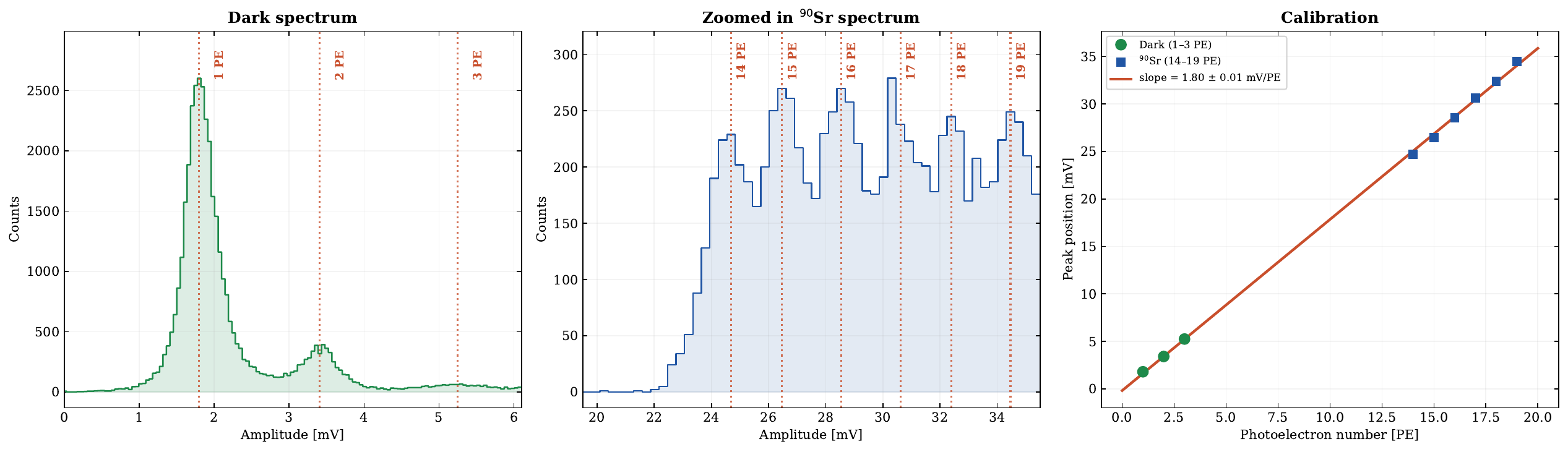}
  \caption{Photoelectron calibration of one readout channel using the dark 
  spectrum (1--3 PE) and a $^{90}$Sr source (higher-order photoelectron peaks), with the 
  combined linear gain fit.} 
  \label{fig:calibration}
\end{figure}

The absolute photoelectron (PE) scale for the deployment configuration with the 560 pF shaping capacitance was determined with an oscilloscope (Tektronix MSO6 series) at \SI{56}{V} bias and room temperature. 
Despite the pulse shaping, the single-photoelectron peaks remain clearly 
resolved: the dark spectrum resolves the 1--3 PE peaks and anchors the absolute scale, while the $^{90}$Sr spectrum resolves the 
higher-order peaks (Figure~\ref{fig:calibration}). A combined linear fit of 
peak position in mV versus photoelectron (PE) number yields an amplitude-to-photoelectron conversion factor of $\SI{1.80}{} ~\pm ~\SI{0.01}{\milli\volt\per PE}$, with an offset consistent with zero. 
An effective crosstalk probability of $\SI{8.1}{} \pm \SI{0.7}{\percent}$ was estimated from the ratio of the averaged dark-count-rate plateaus, measured by a threshold-voltage scan, taking the rate around 1.5\,PE ($\SI{15.3}{} \pm \SI{0.7}{\kilo\hertz}$) to that at 0.5\,PE ($\SI{188.1}{} \pm \SI{12.8}{\kilo\hertz}$). 
The measured rate includes prompt and delayed crosstalk as well as after-pulsing and RF noise (around the 0.5 PE level), and therefore represents an upper bound on the actual prompt 
cross-talk. A signal uniformity study using a collimated $^{90}$Sr source revealed a \SIrange{2}{3}{\percent} shift of the amplitude maximum and the spectrum endpoint by measuring the same readout channel with the source at different positions --- adjacent to the readout WLS fibre and on the opposite face of the cube.

The inter-cube optical crosstalk was assessed on the prototype using a 
${}^{90}$Sr source on a selected cube and reading out all non-irradiated 
neighbouring channels. In the signal region the leakage into a directly adjacent 
row or column remains at the percent level, and 
drops to the few per mille level for non direct adjacent channels.

\begin{figure}[h!]
  \centering
  \includegraphics[width=0.75\linewidth]{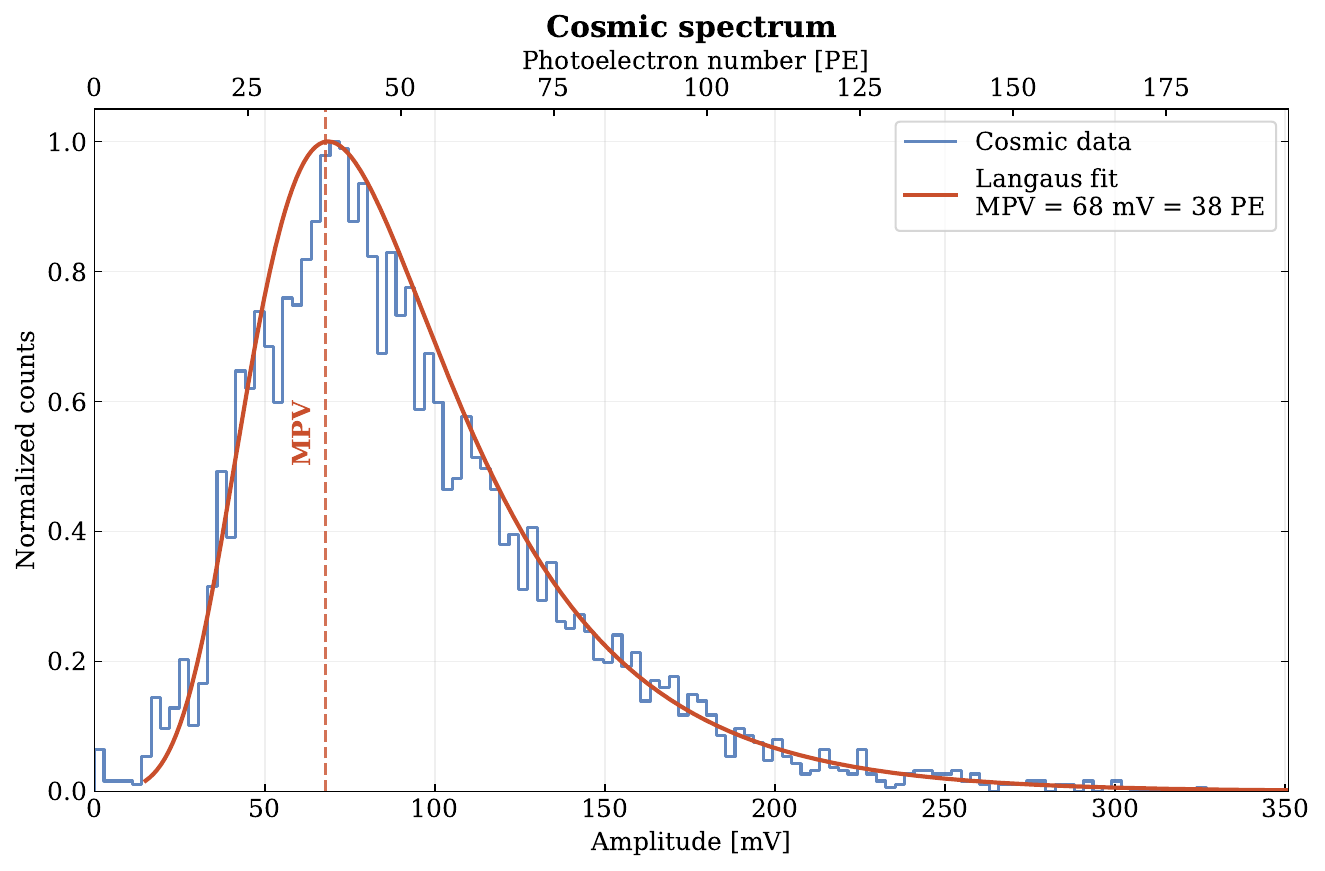}
  \caption{Cosmic-ray amplitude spectrum measured with the oscilloscope using a logical AND 
coincidence of two channels (blue histogram), with a Gaussian-convolved Landau 
(Langaus) fit to the amplitude spectrum (orange curve) overlaid for most probable value determination.}
  \label{fig:cosmic_mV_pe}
\end{figure}

A cosmic amplitude spectrum yields a most probable value (MPV) of 
\SI{68}{\milli\volt}, corresponding to 38\,PE, as indicated by the dashed 
vertical line in Figure~\ref{fig:cosmic_mV_pe}. This sets an upper limit on the 
per-channel discriminator threshold required to detect minimum-ionising 
particles (MIPs). The measured light yield is comparable to the $\sim$40\,PE per 
MIP reported for the SuperFGD detector concept by Abe et al.~\cite{abe2025characterization}.

\subsection{Time-over-Threshold Characterisation}
\label{subsec:tot}

The signal processing chain translates fast SiPM charge pulses into a digital square wave whose duration encodes the deposited energy, enabling direct time-over-threshold (ToT) readout by the FPGA-based data acquisition system. The \SI{560}{\pico\farad} shaping capacitance (see Figure \ref{fig:preamp}) was selected as the base configuration, providing extended dynamic range suited to the expected signal amplitudes in the ROMEO orbit environment.

The RC shaping network at the preamplifier input (\SI{51}{\ohm} in parallel with the shaping capacitance see Figure \ref{fig:preamp}) produces a theoretical nominal time constant of $\tau_{\text{RC}} = \SI{28.56}{\nano\second}$ (\SI{560}{\pico\farad}). Laboratory source measurements yield analogue pulse widths (\SI{10}{\percent} -- \SI{90}{\percent} rise and fall time) of \SI{27.94}{} $\pm ~0.38$~ns which agrees well with the nominal expected value.

Discrimination is performed by a MAX964 ultra-high-speed comparator (propagation delay $t_{\text{pd}} \approx \SI{4.5}{\nano\second}$) \cite{ref:max964} followed by a 74LV14 Schmitt trigger \cite{ref:Schmitt} to ensure sharply defined logic transitions and to prevent high-frequency oscillation of the output for SiPM signals with amplitudes close to the threshold. The resulting digital pulse width follows a logarithmic relationship \cite{reese2026simple,zengin2025embedded,sharma2020estimating,ota2019dual}:
\begin{equation}
\text{ToT} = \tau_{\text{eff}} \cdot \ln\!\left(\frac{A}{V_{\text{thr}}}\right) + t_{\text{offset}}
\label{eq:tot_model}
\end{equation}
where $A$ is the signal amplitude, $V_{\text{thr}}$ is the discriminator threshold adjustable in \SI{0.61}{mV} steps (\SI{2.5}{\volt} reference and 12-bit resolution of the employed DAC), $\tau_{\text{eff}}$ the effective pulse time constant directly determined by the analogue pulse width measurement, and $t_{\text{offset}}$ accounts for propagation delays imposed by the components in the signal chain (e.g.~caused by the employed comparator).

Note that Eq. \ref{eq:tot_model} is applicable in units of either mV or PE, as long as the amplitude value $A$ and the discriminator threshold  $V_{\text{thr}}$ are expressed in the same unit. Working in PE therefore requires converting the threshold voltage using the measured calibration factor (mV per PE) as presented in Section~\ref{subsec:calib}.

The \SI{560}{\pico\farad} shaping capacitor was selected based on the larger $\tau_{\text{eff}}$ of \SI{27.94}{\nano\second}, which leads to higher ToT values and increased sensitivity to amplitude variations. The RC integration inherently suppresses photoelectron-level fluctuations and photon arrival time statistics, whereas the larger time constant further smoothens residual amplitude variations, resulting in more stable ToT measurements.
The sensitivity
\begin{equation}
\frac{\mathrm{d(ToT)}}{\mathrm{d}A}=\frac{\tau_{\text{eff}}}{A}
\end{equation}
is directly proportional to the effective time constant suggesting that a larger shaping capacitance could further improve amplitude resolution, at the cost of reduced signal amplitude and increased pulse length and hence the risk of event pile-up at high event rates.

\begin{figure}[h]
    \centering
    \includegraphics[width=\linewidth]{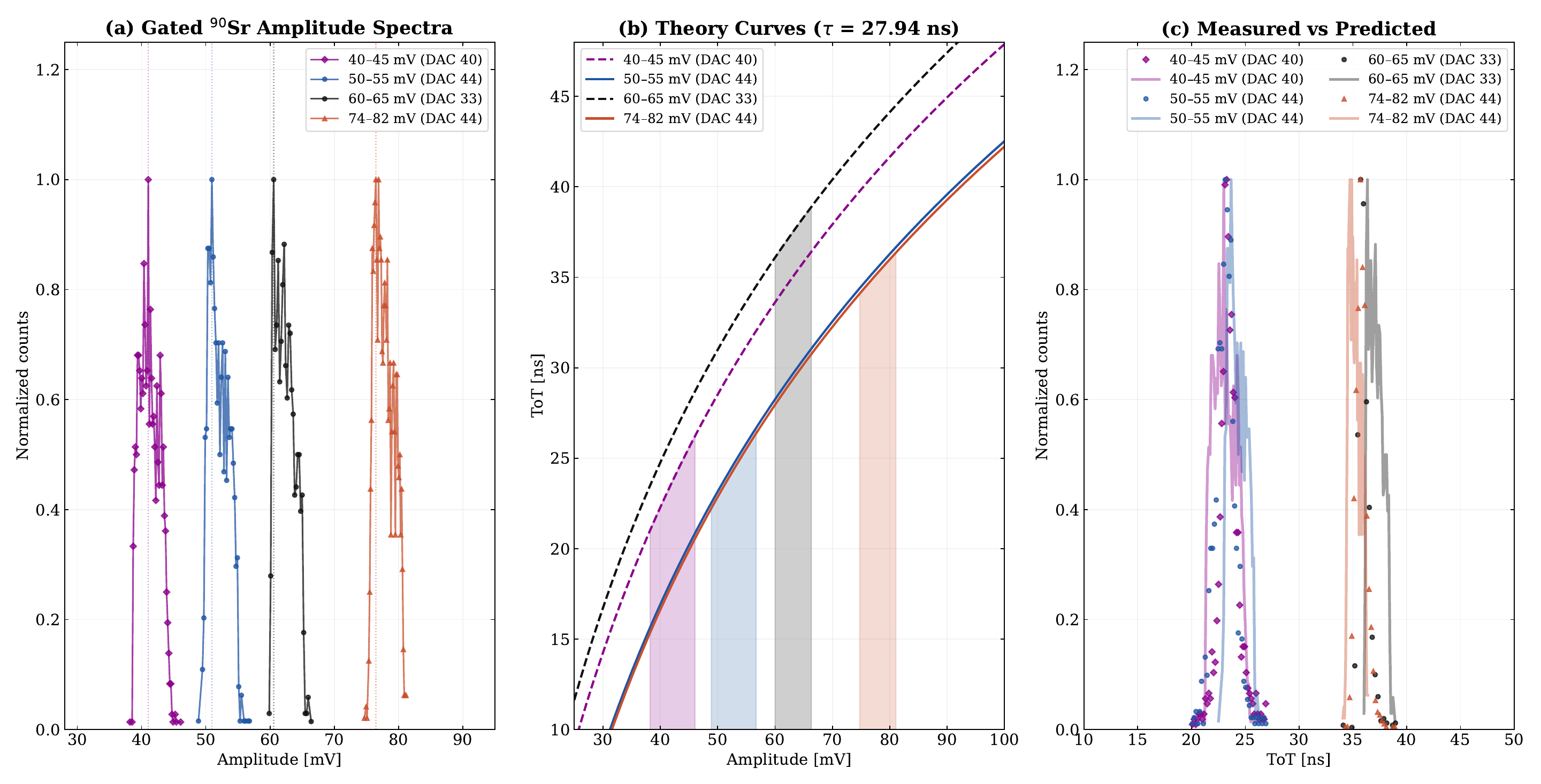}
    \caption{Comparison between measured comparator pulse length and model-predicted ToT distributions from gated $^{90}$Sr measurements and \SI{560}{\pico\farad} shaping capacitance (see text). The shaded bands indicate the amplitude gate used for evaluation. Model parameters: $\tau_{\text{eff}} = \SI{27.94}{\nano\second}$, $V_{\text{thr}}=\SI{20.14}{mV}$ (DAC 33), $\SI{24.41}{mV}$ (DAC 40), $\SI{26.86}{mV}$ (DAC 44). An overdrive-dependent walk $t_{\text{offset}}= \SIrange{8}{4}{ns}$ was taken into account as specified in the data-sheet for the employed comparator device~\cite{ref:dac}.}
  \label{fig:tot}
\end{figure}

Model validation was performed using a $^{90}$Sr source. A \SI{50}{\ohm} splitter was employed to simultaneously acquire the analogue signal with a fast sampling oscilloscope (Tektronix MSO6 series) and the discriminator digital output for the same recorded event. The oscilloscope trigger was set on the analogue channel with an amplitude acceptance gate constraining the signal to, e.g.~\SI{50}{} -- \SI{55}{\milli\volt}, ensuring that the analogue amplitude and digital ToT were recorded synchronously for the same gated event population, allowing a direct comparison between the measured and model-predicted ToT distributions as shown in Figure~\ref{fig:tot}.

\begin{figure}[h!]
    \centering
    \includegraphics[width=0.65\linewidth]{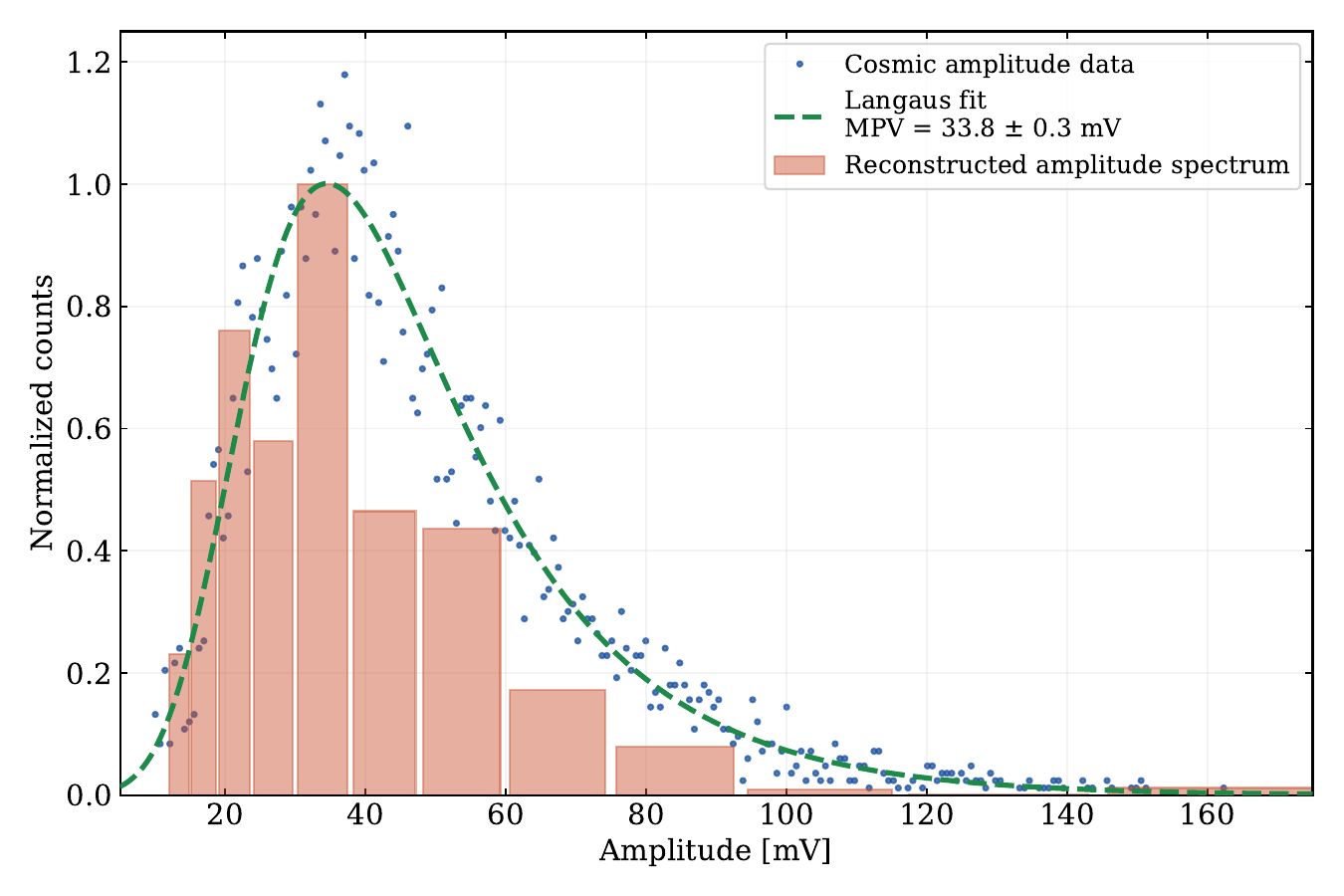}
    \caption{Amplitude spectrum reconstructed from the measured digital ToT distribution using the inverse model (\ref{eq:tot_model_amp}) with a nominal offset of \SI{4}{\nano\second} and $V_{\text{thr}} = \SI{10.38}{\milli\volt}$ (red histogram bars). Uniform ToT bins of \SI{6}{\nano\second} map to exponentially wider amplitude intervals at higher absolute amplitudes due to the logarithmic nature of the ToT--amplitude conversion. For the same measurement, the amplitude spectrum captured at the comparator input (blue points) and the fit consisting of a Gaussian-convolved Landau distribution to these data points (green curve) is overlaid for comparison.}
  \label{fig:cosmic}
\end{figure}

The propagation delay of the MAX964 comparator is inversely dependent on the overdrive voltage, ranging from \SI{4}{\nano\second} (typical, high overdrive) to \SI{8}{\nano\second} (low overdrive)~\cite{ref:max964}. In the selected configuration ($V_\mathrm{thr} = \SI{10}{\milli\volt}$), the overdrive varies across the amplitude spectrum, but at the most probable value of the cosmic energy loss distribution of $\approx\SI{34}{\milli\volt}$ shown in Figure~\ref{fig:cosmic}, it amounts to only $\approx\SI{24}{\milli\volt}$.

The comparator's propagation delay sets a fundamental limit on the ToT resolution: with a best-case value of $t_\mathrm{pd} \approx \SI{4}{\nano\second}$ at high overdrive, the ToT spectrum can be binned no finer than \SI{4}{\nano\second}, increasing toward $\approx\SI{8}{\nano\second}$ at low overdrive near threshold. This sets the achievable bin width for the amplitude reconstruction and translates directly into a relative amplitude resolution of $\Delta A/A = t_\mathrm{pd}/\tau_\mathrm{eff}$, ranging from approximately \SI{14}{\percent} to \SI{29}{\percent}. This is sufficient to discriminate between minimum-ionising particles and stopped protons, whose energy deposits differ significantly.

The amplitude spectrum can equivalently be reconstructed from the measured ToT distribution by inverting Eq.~\eqref{eq:tot_model}:
\begin{equation}
A = V_{\text{thr}} \exp[(\text{ToT} - t_{\text{offset}})/\tau_{\text{eff}}]
\label{eq:tot_model_amp}
\end{equation}
Applying this inverse transformation to the measured cosmic ToT spectrum, binned at \SI{6}{\nano\second} yields the 
reconstructed amplitude distribution shown in Figure~\ref{fig:cosmic}. The reconstruction reproduces the expected Landau-like shape of the cosmic-ray amplitude spectrum, with the most probable value (MPV) of \SI{33.8}{\milli\volt} in agreement with the Langaus (Landau convolved with Gaussian) fit of the directly measured analogue amplitude spectrum. 

Note that the amplitude spectrum was acquired through a resistive \SI{50}{\ohm} splitter, which halves the signal amplitude at each output; the measured MPV of \SI{33.8}{\milli\volt} therefore corresponds to a true detector amplitude of approximately \SI{68}{\milli\volt}, or approximately \SI{38}{PE} using the amplitude-to-photoelectron conversion factor of Section~\ref{subsec:calib}, consistent with the finding of the cosmic measurement in Figure ~\ref{fig:cosmic_mV_pe}.

\subsection{Proton Beam Test: First Track Reconstruction}
\label{subsec:tracks}

\begin{figure}[h]
  \begin{subfigure}[b]{0.5\textwidth}
    \centering
    \includegraphics[width=\linewidth]{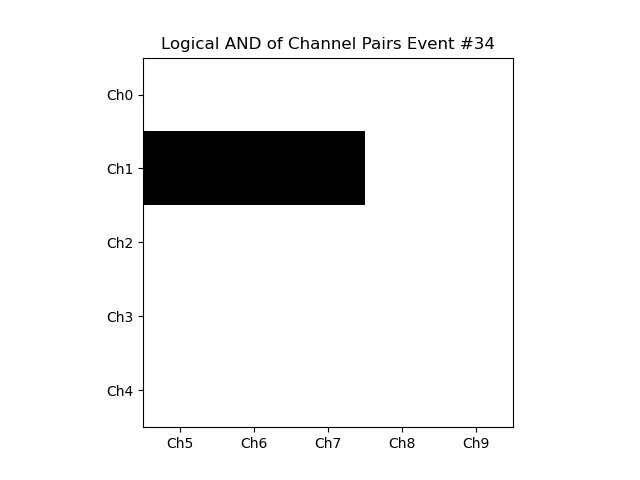}
  \end{subfigure}
  \hfill
  \begin{subfigure}[b]{0.5\textwidth}
    \centering
    \includegraphics[width=\linewidth]{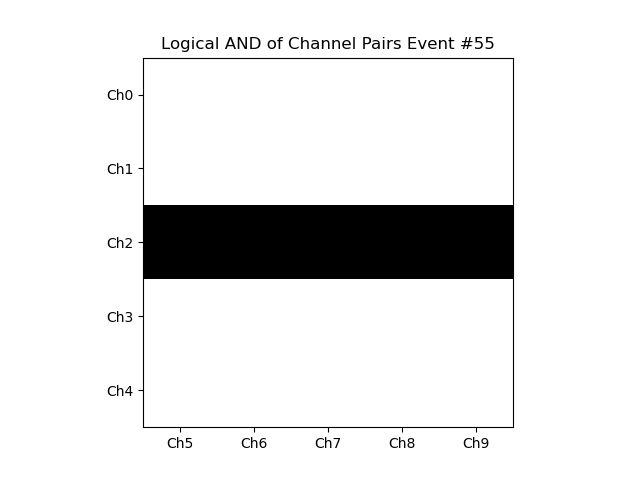}
  \end{subfigure}
  \caption{Representative hit patterns recorded during the \SI{220}{\MeV} proton beam test at the Marburg Ion Beam Therapy Center (MIT). Spatially correlated tracks with orientations consistent with the expected  scattering geometry are clearly visible in the $5{\times}5$ channel matrix.}
  \label{fig:tracks}
\end{figure}
A beam test was performed at the Marburg Ion Beam Therapy Center (MIT, Marburg, Germany) \cite{zink2024organization} using \SI{220}{\MeV} protons with an intensity of \SI{6.9e8}{protons \per \second} with a beam spot size of \SI{9.8}{\milli\meter} \cite{witt2026feasibility}.
The prototype was deliberately placed off-axis relative to the primary beam behind a target consisting of electronic components mounted on a standard FR4 PCB that was being irradiated. The detector plane was tilted by approximately \SIrange{5}{10}{\degree} to the beam axis, so that scattered protons would traverse the \(5{\times}5\) sensitive layer close to normal incidence.
This geometry was chosen to test whether the hit-pattern readout can resolve the spatial structure of straight particle tracks --- a fundamental requirement for directional reconstruction in the flight instrument.

In the hit patterns recorded, spatially correlated tracks as line-like structures were clearly identified in the $5{\times}5$ channel matrix, with orientations consistent with the expected scattering geometry
(Figure~\ref{fig:tracks}). While the angular precision achievable with a single \(5{\times}5\) layer is limited and a quantitative track fit is beyond the scope of this prototype test, the observation of directionally correlated hit patterns constitutes a direct proof of concept for the 3D track reconstruction capability of the full \(5{\times}5{\times}5\) instrument. In addition, the data provide a first indication that rate asymmetries between channels carry information about the dominant particle direction --- an important observable for the orbit-wide rate-mapping objective of RUBIK.

\section{Summary}
\label{sec:summary}

We have presented RUBIK, a \(5{\times}5{\times}5\) PVT-based scintillator cube array with 50 WLS-fibre and SiPM readout per channel. It features custom analogue electronics incorporating commercial off-the-shelf philosophy with per-channel programmable thresholds, an FPGA-based data acquisition system, and a dedicated power and bias distribution board, developed as the scientific particle detector payload for the ROMEO satellite mission.

A \(5{\times}5{\times}1\) single-layer prototype has validated the detector concept in three key areas. The SiPM response was characterised as a function of bias voltage and temperature, establishing the in-orbit gain-stabilisation strategy. The time-over-threshold amplitude 
reconstruction method was validated against radioactive source and cosmic ray measurements, confirming good agreement between the directly measured and ToT-reconstructed amplitude spectra. A proton beam test at \SI{220}{\MeV} showed spatially correlated hit-pattern clusters consistent with the expected scattering geometry, confirming that directional information is encoded in the single-layer readout and providing a proof of concept for 3D track reconstruction in the full instrument.

SRIM-2013 range simulations of the complete detector stack establish three distinct interaction regimes that define the science acceptance window. Protons below approximately \SI{80}{\MeV} stop before reaching the most distal cube layer and can be rejected by the five-fold coincidence requirement. Protons in the \SIrange{80}{100}{\MeV} range traverse all five layers but stop within the fifth cube, producing a Bragg-peak enhancement in the time-over-threshold signal of the last layer that distinguishes them from traversing particles. At energies beyond \SI{100}{\MeV}, protons traverse the full detector stack with an approximately flat ionisation profile. Just above the \SI{100}{\MeV} threshold the per-cube deposit is around \SI{3.8}{\MeV}, decreasing towards the 
minimum-ionising value of approximately $\SI{2}{\MeV}$ per cube at higher energies. This 
flat, penetrating topology is the primary expected signal for GCR detection. The combination of five-fold coincidence and per-cube ToT therefore provides an effective science acceptance threshold of approximately \SI{100}{\MeV} without a dedicated energy measurement stage.

RUBIK will target two science objectives enabled uniquely by the
ROMEO walking-orbit profile: a spatial map of charged-particle rates
across the full altitude range from \SI{600}{km} to the inner radiation
belt, and the first exploration of event-by-event ground-to-space
correlation between primary cosmic-ray events detected in orbit and
secondary air showers observed at ground level. In addition, the per cube Time-over-Threshold readout provides a coarse 
energy-deposition measurement along each track, adding calorimetric information 
to the rate and directional data.

\section*{Acknowledgements}

The authors thank the ROMEO team of IRS Stuttgart for the opportunity to contribute to the mission and for their continued support. Special thanks go to the IRS group leader Prof. Dr.-Ing. Sabine Klinkner and to Thorben Löffler. The authors also appreciate the technical support provided by the in-house engineers Ren{\'e} Schubert and Christian Schinz.

\bibliography{biblio.bib}
\end{document}

%% file: preamp_rubik.tex
\begin{tikzpicture}[transform shape]
	\node[plain mono amp, xscale=1.15, yscale=1.15] at (6.431, 6){};
	\draw (2.25, 7.75) to[empty diode, /tikz/circuitikz/bipoles/length=1.75cm, l_={$\mathrm{S13360 - 1350PE}$}] (2.25, 8.75);
	\draw (3.75, 6) to[capacitor, l={$\mathrm{560\, pF}$}, label distance=-0.02cm] (3.75, 4.5);
	\draw (5, 6) to[capacitor, l_={$\mathrm{100 \,nF}$}] (3.75, 6);
	\draw (7.859, 6) to[capacitor, l={$\mathrm{100 \,nF}$}] (8.729, 6);
	\node[ground] at (3, 4.5){};
	\draw (2.25, 6) to[american resistor, l_={$\mathrm{51 \,\Omega}$}] (2.25, 4.5);
	\draw (2.5, 4.5) -- (3.75, 4.5);
	\node[msport, xscale=1.75, yscale=1.75](N1) at (2.775, 9.75){} node[anchor=center] at (3.375, 9.75){$\mathrm{V_{Bias}}$};
	\node[msport, xscale=-1.75, yscale=-1.75](N2) at (10.454, 6){} node[anchor=center] at (9.754, 6){$\mathrm{Out}$};
	\draw (8.729, 6) -- (9.229, 6);
	\draw (3.75, 6) -- (2.5, 6);
	\draw (2.75, 9.75) -- (2.25, 9.75) -| (2.25, 8.75);
	\draw (2.25, 7.75) -| (2.25, 6);
	\draw (2.5, 4.5) -- (2.25, 4.5);
	\draw (2.5, 6) -- (2.25, 6);
	\node[shape=rectangle, minimum width=1.965cm, minimum height=0.465cm] at (6.314, 6.097){} node[anchor=north west, align=left, text width=1.577cm, inner sep=6pt] at (5.314, 6.347){BGA2815};
	\draw (5.063, 6) |- (5, 6);
	\draw (7.859, 6) -- (7.8, 6);
	\node[circ] at (2.25, 6){};
	\node[circ] at (3.75, 6){};
	\node[circ] at (3, 4.5){};
\end{tikzpicture}

%% file: biblio.bib
@inproceedings{Loeffler2022,
  title={Preliminary design of the radiation protection of the romeo satellite in the lower medium earth orbit},
  author={Loeffler, T and Burgdorf, Jonas and Hildebrandt, J{\'e}r{\^o}me and Boetsch-Zav{\v{r}}el, Lena and Holeczek, Cedric and Paetschke, Susann and Kanzow, Marlin and Largent, Patrick and Petri, Jona and Lengowski, Michael and others},
  booktitle={73rd International Astronautical Congress, Paris, France},
  year={2022}
}

@article{baker2018space,
  title={Space weather effects in the Earth’s radiation belts},
  author={Baker, DN and Erickson, PJ and Fennell, JF and Foster, JC and Jaynes, AN and Verronen, PT},
  journal={Space Science Reviews},
  volume={214},
  number={1},
  pages={17},
  year={2018},
  publisher={Springer}
}

@article{khrenov2017first,
  title={First results from the TUS orbital detector in the extensive air shower mode},
  author={Khrenov, BA and Klimov, PA and Panasyuk, MI and Sharakin, SA and Tkachev, LG and Zotov, M Yu and Biktemerova, SV and Botvinko, AA and Chirskaya, NP and Eremeev, VE and others},
  journal={Journal of Cosmology and Astroparticle Physics},
  volume={2017},
  number={09},
  pages={006--006},
  year={2017}
}

@article{zheng2024overview,
  title={Overview, progress and next steps for our understanding of the near-earth space radiation and plasma environment: Science and applications},
  author={Zheng, Yihua and Jun, Insoo and Tu, Weichao and Shprits, Yuri Y and Kim, Wousik and Matthi{\"a}, Daniel and Meier, Matthias M and Tobiska, W Kent and Miyoshi, Yoshizumi and Jordanova, Vania K and others},
  journal={Advances in Space Research},
  year={2024},
  publisher={Elsevier}
}

@article{barmin2014near,
  title={Near-Earth space as an object of global monitoring},
  author={Barmin, IV and Kulagin, VP and Savinykh, VP and Tsvetkov, V Ya},
  journal={Solar System Research},
  volume={48},
  number={7},
  pages={531--535},
  year={2014},
  publisher={Springer}
}

@article{spence2022achievements,
  title={Achievements and lessons learned from successful small satellite missions for space weather-oriented research},
  author={Spence, Harlan E and Caspi, Amir and Bahcivan, Hasan and Nieves-Chinchilla, Jesus and Crowley, Geoff and Cutler, James and Fish, Chad and Jackson, David and Jorgensen, Therese Moretto and Klumpar, David and others},
  journal={Space Weather},
  volume={20},
  number={7},
  pages={e2021SW003031},
  year={2022},
  publisher={Wiley Online Library}
}

@article{adriani2011pamela,
  title={PAMELA measurements of cosmic-ray proton and helium spectra},
  author={Adriani, Oea and Barbarino, GC and Bazilevskaya, GA and Bellotti, Roberto and Boezio, M and Bogomolov, EA and Bonechi, Lorenzo and Bongi, Massimo and Bonvicini, V and Borisov, S and others},
  journal={Science},
  volume={332},
  number={6025},
  pages={69--72},
  year={2011},
  publisher={American Association for the Advancement of Science}
}

@article{AMS2015,
  author  = {Aguilar, M. and others (AMS Collaboration)},
  title   = {Precision Measurement of the Proton Flux in Primary Cosmic
             Rays from Rigidity 1 {GV} to 1.8 {TV} with the {Alpha Magnetic
             Spectrometer} on the {International Space Station}},
  journal = {Phys.\ Rev.\ Lett.},
  volume  = {114},
  pages   = {171103},
  year    = {2015},
  doi     = {10.1103/PhysRevLett.114.171103}
}

@article{CALET2019,
  author  = {Adriani, O. and others (CALET Collaboration)},
  title   = {Direct Measurement of the Cosmic-Ray Proton Spectrum from
             50 {GeV} to 10 {TeV} with the {Calorimetric Electron Telescope}
             on the {International Space Station}},
  journal = {Phys.\ Rev.\ Lett.},
  volume  = {122},
  pages   = {181102},
  year    = {2019},
  doi     = {10.1103/PhysRevLett.122.181102}
}

@article{DAMPE2019,
  author  = {An, Q. and others (DAMPE Collaboration)},
  title   = {Measurement of the cosmic ray proton spectrum from 40 {GeV}
             to 100 {TeV} with the {DAMPE} satellite},
  journal = {Sci.\ Adv.},
  volume  = {5},
  pages   = {eaax3793},
  year    = {2019},
  doi     = {10.1126/sciadv.aax3793}
}

@article{JEMEUSO2020,
  author  = {Adams, J.~H. and others (JEM-EUSO Collaboration)},
  title   = {The {JEM-EUSO} mission: {A} review of its scientific
             objectives and instrumentation},
  journal = {Exp.\ Astron.},
  volume  = {40},
  pages   = {281},
  year    = {2015},
  doi     = {10.1007/s10686-015-9482-x}
}

@article{ziegler2010,
  title={{SRIM--The stopping and range of ions in matter (2010)}},
  author={Ziegler, James F and Ziegler, Matthias D and Biersack, Jochen P},
  journal={Nuclear Instruments and Methods in Physics Research Section B: Beam Interactions with Materials and Atoms},
  volume={268},
  number={11-12},
  pages={1818--1823},
  year={2010},
  publisher={Elsevier}
}

@article{IceCube2013,
  author  = {Aartsen, M.~G. and others (IceCube Collaboration)},
  title   = {Observation of High-Energy Astrophysical Neutrinos in
             Three Years of {IceCube} Data},
  journal = {Phys.\ Rev.\ Lett.},
  volume  = {113},
  pages   = {101101},
  year    = {2014},
  doi     = {10.1103/PhysRevLett.113.101101}
}

@article{Auger2015,
  author  = {Aab, A. and others (Pierre Auger Collaboration)},
  title   = {The {Pierre Auger} Observatory: Contributions to the
             34th International Cosmic Ray Conference ({ICRC2015})},
  journal = {arXiv:1509.03732},
  year    = {2015}
}

@article{TA2013,
  author  = {Abu-Zayyad, T. and others (Telescope Array Collaboration)},
  title   = {The surface detector array of the {Telescope Array}
             experiment},
  journal = {Nucl.\ Instrum.\ Meth.\ A},
  volume  = {689},
  pages   = {87},
  year    = {2012},
  doi     = {10.1016/j.nima.2012.05.079}
}

@techreport{ap8,
  title={{AP-8 trapped proton environment for solar maximum and solar minimum}},
  author={Sawyer, Donald M and Vette, James I},
  year={1976}
}

@inproceedings{ap8ae8,
  title={{Models of the Trapped Radiation Environment. NASA SP-3024}},
  author={James I. Vette and Antonio B. Lucero and Jon A. Wright and Joseph Herbert King and James P. Lavine and William L. Imhof and C. O. Bostrom and David S. Beall and John Charles Armstrong and Harry H. Heckman and Peter J. Lindstrom and George H. Nakano and George A. Paulikas and J. Bernard Blake},
  year={1966},
}

@inproceedings{kruglanski2009space,
  title={{Space environment information system (SPENVIS)}},
  author={Kruglanski, Michel and Messios, Neophytos and De Donder, Erwin and Gamby, Emmanuel and Calders, Stijn and Hetey, Laszlo and Evans, Hugh},
  booktitle={EGU General Assembly Conference Abstracts},
  pages={7457},
  year={2009}
}

@article{heynderickx2004new,
  title={New radiation environment and effects models in the {European Space Agency's Space Environment Information System (SPENVIS)}},
  author={Heynderickx, D and Quaghebeur, B and Wera, J and Daly, EJ and Evans, HDR},
  journal={Space Weather},
  volume={2},
  number={10},
  year={2004},
  publisher={Wiley Online Library}
}

@inproceedings{kruglanski2009last_spenvis,
  title={Last upgrades and development of the space environment information system (SPENVIS)},
  author={Kruglanski, Michel and Messios, Neophytos and De Donder, Erwin and Gamby, Emmanuel and Calders, Stijn and Hetey, Laszlo and Evans, Hugh and Daly, Eamonn},
  booktitle={2009 European Conference on Radiation and Its Effects on Components and Systems},
  pages={563--565},
  year={2009},
  organization={IEEE}
}

@article{ginet2013ae9,
  title={{AE9, AP9 and SPM: New models for specifying the trapped energetic particle and space plasma environment}},
  author={Ginet, GP and O’Brien, TP and Huston, SL and Johnston, WR and Guild, TB and Friedel, R and Lindstrom, CD and Roth, CJ and Whelan, P and Quinn, RA and others},
  journal={Space science reviews},
  volume={179},
  number={1},
  pages={579--615},
  year={2013},
  publisher={Springer}
}

@article{blondel2018fully,
  title={A fully-active fine-grained detector with three readout views},
  author={Blondel, A and Cadoux, F and Fedotov, S and Khabibullin, M and Khotjantsev, A and Korzenev, A and Kostin, A and Kudenko, Y and Longhin, A and Mefodiev, A and others},
  journal={Journal of Instrumentation},
  volume={13},
  number={02},
  pages={P02006--P02006},
  year={2018}
}

@article{ching2021ultrahigh,
  title={Ultrahigh-speed point scanning two-photon microscopy using high dynamic range silicon photomultipliers},
  author={Ching-Roa, Vincent D and Olson, Eben M and Ibrahim, Sherrif F and Torres, Richard and Giacomelli, Michael G},
  journal={Scientific Reports},
  volume={11},
  number={1},
  pages={5248},
  year={2021},
  publisher={Nature Publishing Group UK London}
}

@misc{ref:biaspcb_github,
title={{OpenSiPM GitHub repository}},
howpublished={\url{https://github.com/OpenSiPM/sipm-bias-control/tree/master/bias-control-lt8362}} 
}

@article{xapsos1999probability,
  title={Probability model for worst case solar proton event fluences},
  author={Xapsos, MA and Summers, GP and Barth, JL and Stassinopoulos, EG and Burke, EA},
  journal={IEEE Transactions on Nuclear Science},
  volume={46},
  number={6},
  pages={1481--1485},
  year={1999},
  publisher={IEEE}
}

@article{xapsos2000probability,
  title={Probability model for cumulative solar proton event fluences},
  author={Xapsos, MA and Summers, GP and Barth, JL and Stassinopoulos, EG and Burke, EA},
  journal={IEEE Transactions on Nuclear Science},
  volume={47},
  number={3},
  pages={486--490},
  year={2000},
  publisher={IEEE}
}

@article{xapsos2007model,
  title={Model for cumulative solar heavy ion energy and linear energy transfer spectra},
  author={Xapsos, MA and Stauffer, C and Jordan, T and Barth, JL and Mewaldt, RA},
  journal={IEEE Transactions on Nuclear Science},
  volume={54},
  number={6},
  pages={1985--1989},
  year={2007},
  publisher={IEEE}
}

@misc{iso2004space,
  title={Space environment (natural and artifical)-galactic cosmic ray model},
  author={ISO, I},
  year={2004},
  publisher={Geneva, Switzerland: International Organization for Standardization}
}

@article{blondel2020superfgd,
  title={{The SuperFGD prototype charged particle beam tests}},
  author={Blondel, Alain and Bogomilov, M and Bordoni, S and Cadoux, F and Douqa, D and Dugas, K and Ekelof, T and Favre, Y and Fedotov, S and Fransson, Kjell and others},
  journal={Journal of Instrumentation},
  volume={15},
  number={12},
  pages={P12003--P12003},
  year={2020}
}

@article{mineev2019beam,
  title={{Beam test results of 3D fine-grained scintillator detector prototype for a T2K ND280 neutrino active target}},
  author={Mineev, O and Blondel, A and Favre, Y and Fedotov, S and Khotjantsev, A and Korzenev, A and Kudenko, Yu and Mefodiev, A and Noah, E and Sgalaberna, D and others},
  journal={Nuclear Instruments and Methods in Physics Research Section A: Accelerators, Spectrometers, Detectors and Associated Equipment},
  volume={923},
  pages={134--138},
  year={2019},
  publisher={Elsevier}
}

@misc{ref:df-2000,
title={{DF-2000MA} datasheet},
howpublished={\url{https://multimedia.3m.com/mws/media/982449O/3mtm-specular-film-df2000ma-technical-data-sheet.pdf}} 
}

@misc{ref:ej-248,
title={{EJ-248} datasheet},
howpublished={\url{https://eljentechnology.com/products/plastic-scintillators/ej-244-ej-248-ej-244m-ej-248m}} 
}

@misc{ref:epoxy,
title={{EPO-TEK}{®} 301-2 data sheet},
howpublished={\url{https://www.matweb.com/search/datasheet.aspx?matguid=6d77c56de39a414eb47a3a3b6856b02a&ckck=1}} 
}

@misc{ref:wls,
title={Y-11 data sheet},
howpublished={\url{https://methacrylate.kuraray.com/content/dam/methacrylate/en/documents/psf.pdf}} 
}

@misc{ref:sipm,
title={{S13360-1350PE} data sheet},
howpublished={\url{https://www.hamamatsu.com/eu/en/product/optical-sensors/mppc/mppc_mppc-array/S13360-1350PE.html}} 
}

@article{abe2025characterization,
  title={Characterization of the optical model of the T2K 3D segmented plastic scintillator detector unit cube},
  author={Abe, Seisho and Alekseev, Igor and Arai, T and Arihara, T and Arimoto, S and Babu, Nayan and Baranov, Vladimir and Bartoszek, L and Berns, Lukas and Bhattacharjee, S and others},
  journal={Nuclear Instruments and Methods in Physics Research Section A: Accelerators, Spectrometers, Detectors and Associated Equipment},
  volume={1080},
  pages={170757},
  year={2025},
  publisher={Elsevier}
}

@misc{ref:tmp235,
title={{TMP235} data sheet},
howpublished={\url{https://www.ti.com/product/de-de/TMP235}} 
}

@misc{ref:ltc2986,
title={{LTC2986} data sheet},
howpublished={\url{https://www.analog.com/en/products/ltc2986.html}} 
}

@misc{ref:bga2800,
title={{BGA2800} data sheet},
howpublished={\url{https://www.nxp.com/docs/en/data-sheet/BGA2800.pdf}} 
}

@misc{ref:bga2815,
title={{BGA2815} data sheet},
howpublished={\url{https://www.nxp.com/docs/en/data-sheet/BGA2815.pdf}} 
}

@misc{ref:dac,
title={{AD5674R} data sheet},
howpublished={\url{https://www.analog.com/en/products/ad5674r.html}} 
}

@misc{ref:fpga,
title={{TE0714} module overview},
howpublished={\url{https://wiki.trenz-electronic.de/display/PD/TE0714+TRM}} 
}

@misc{ref:attiny,
title={{ATtiny816} overview},
howpublished={\url{https://www.microchip.com/en-us/product/attiny816}} 
}

@misc{ref:ccsds,
title={{Consultative Committee for Space Data Systems CCSDS}},
howpublished={\url{https://ccsds.org/}} 
}

@misc{ref:max964,
title={{Ultra-High-Speed Beyond-the-Rails Comparator MAX964}},
howpublished={\url{www.analog.com/en/products/max964.html}} 
}

@article{burmistrov2025performance,
  title={Performance and radiation damage mitigation strategy for silicon photomultipliers on LEO space missions},
  author={Burmistrov, L and Davarpanah, S and Heller, M and Montaruli, T and Trimarelli, C and Aloisio, R and Barbato, FCT and De Mitri, I and Di Giovanni, A and Fontanella, G and others},
  journal={Journal of Cosmology and Astroparticle Physics},
  volume={2025},
  number={07},
  pages={073},
  year={2025},
  publisher={IOP Publishing}
}

@article{zheng2022orbit,
  title={In-orbit radiation damage characterization of SiPMs in the GRID-02 CubeSat detector},
  author={Zheng, Xutao and Gao, Huaizhong and Wen, Jiaxing and Zeng, Ming and Pan, Xiaofan and Xu, Dacheng and Liu, Yihui and Zhang, Yuchong and Peng, Haowei and Jiang, Yuchen and others},
  journal={Nuclear Instruments and Methods in Physics Research Section A: Accelerators, Spectrometers, Detectors and Associated Equipment},
  volume={1044},
  pages={167510},
  year={2022},
  publisher={Elsevier}
}

@article{garutti2019radiation,
  title={Radiation damage of SiPMs},
  author={Garutti, E and Musienko, Yu},
  journal={Nuclear Instruments and Methods in Physics Research Section A: Accelerators, Spectrometers, Detectors and Associated Equipment},
  volume={926},
  pages={69--84},
  year={2019},
  publisher={Elsevier}
}

@misc{ref:Schmitt,
title={Hex inverting Schmitt trigger},
howpublished={\url{https://www.nexperia.com/products/analog-logic-ics/logic/buffers-inverters-transceivers/schmitt-trigger-ics/series/74LV14.html}} 
}

@misc{ref:lmr33620,
title={LMR33620 data sheet},
howpublished={\url{https://www.ti.com/lit/ds/symlink/lmr33620.pdf}} 
}

@misc{ref:lt8361,
title={LT8361 boost converter},
howpublished={\url{https://www.analog.com/en/products/lt8361.html}} 
}

@misc{ref:ldo,
title={Rt9072B data sheet},
howpublished={\url{https://www.digchip.com/datasheets/parts/datasheet/2/403/RT9072BGB-pdf.php}} 
}

@article{zengin2025embedded,
  title={An Embedded FPGA Design for Double-Time-Over-Threshold (DToT) Measurement},
  author={Zengin, Ayd{\i}n Tar{\i}k},
  journal={Balkan Journal of Electrical and Computer Engineering},
  volume={13},
  number={4},
  pages={382--387},
  year={2025},
  publisher={MUSA YILMAZ}
}

@article{fiedler2026pbwo4,
  title={A PbWO4 scintillator range-telescope for real-time depth-dose measurement in low-dose environments for proton therapy},
  author={Fiedler, Niclas and Kazlou, Dzmitry and Dormenev, Valerii and Zaunick, Hans-Georg and Derksen, Larissa and Brinkmann, Kai-Thomas and Baumann, Kilian-Simon},
  journal={Journal of Instrumentation},
  volume={21},
  number={05},
  pages={P05018},
  year={2026},
  publisher={IOP Publishing}
}

@article{kazlou2026red,
  title={Red Pitaya DAQ for fast neutron detection: a scalable, compact and low-cost solution},
  author={Kazlou, Dzmitry and Bergert, Roman and Zaunick, Hans-Georg and Brinkmann, Kai-Thomas},
  journal={Journal of Instrumentation},
  volume={21},
  number={07},
  pages={P07005},
  year={2026},
  publisher={IOP Publishing}
}

@article{zink2024organization,
  title={Organization and operation of multi particle therapy facilities: the Marburg Ion-Beam Therapy Center, Germany (MIT)},
  author={Zink, Klemens and Baumann, Kilian Simon and Theiss, Ulrike and Subtil, Florentine and Lahrmann, Sonja and Eberle, Fabian and Adeberg, Sebastian},
  journal={Health and Technology},
  volume={14},
  number={5},
  pages={929--938},
  year={2024},
  publisher={Springer}
}

@article{witt2026feasibility,
  title={Feasibility of proton daily QA prototype for pencil beam scanning carbon ion beam therapy},
  author={Witt, Matthias and Flatten, Veronika and Sch{\"o}nfeld, Andreas and Baumann, Kilian-Simon and Weber, Uli and Zink, Klemens},
  journal={Journal of Applied Clinical Medical Physics},
  volume={27},
  number={2},
  pages={e70502},
  year={2026},
  publisher={Wiley Online Library}
}

@misc{ref:gecko,
title={Harwin Gecko connector},
howpublished={\url{https://www.harwin.com/products/G125-2241096F5}} 
}

@article{reese2026simple,
  title={A simple dynamic time-over-threshold circuit for high-resolution amplitude measurement},
  author={Reese, Michael},
  journal={Nuclear Instruments and Methods in Physics Research A},
  volume={1083},
  number={17116},
  pages={5},
  year={2026}
}

@article{ota2019dual,
  title={Dual time-over-threshold: estimation of decay time and pulse height for scintillation detectors},
  author={Ota, R and Nakajima, K and Ogawa, I and Tamagawa, Y},
  journal={Journal of Instrumentation},
  volume={14},
  number={11},
  pages={P11012--P11012},
  year={2019}
}

@article{sharma2020estimating,
  title={Estimating relationship between the time over threshold and energy loss by photons in plastic scintillators used in the J-PET scanner},
  author={Sharma, Sushil and Chhokar, Jyoti and Curceanu, Catalina and Czerwi{\'n}ski, Eryk and Dadgar, Meysam and Dulski, Kamil and Gajewski, J and Gajos, Aleksander and Gorgol, M and Gupta-Sharma, Neha and others},
  journal={EJNMMI physics},
  volume={7},
  number={1},
  pages={39},
  year={2020},
  publisher={Springer}
}
